\documentclass[useAMS,usenatbib]{mn2e}
\usepackage{epsf}
\usepackage{amssymb}
\usepackage{graphicx}
\usepackage[usenames]{color}

\newcommand {\bc}{\begin {center}}
\newcommand {\ec}{\end {center}}
\newcommand {\be}{\begin {equation}}
\newcommand {\ee}{\end {equation}}

\def\disp {\displaystyle}

\title[Quantifying properties of ICM inhomogeneities]{Quantifying properties of ICM inhomogeneities}
\author[Zhuravleva et al.]{I. Zhuravleva$^{1}$\thanks{izhur@mpa-garching.mpg.de},
  E. Churazov$^{1,2}$, A. Kravtsov$^{3,4}$, E.T. Lau$^{5,6}$,
  D. Nagai$^{5,6}$, \newauthor R. Sunyaev$^{1,2}$\\ \\
$^{1}$MPI f\"ur Astrophysik, Karl-Schwarzschild str. 1, Garching, 85741, Germany\\
$^{2}$Space Research Institute, Profsoyuznaya str. 84/32, Moscow, 117997, Russia\\
$^3$Department of Astronomy and Astrophysics, University of Chicago,
  5640 South Ellis Avenue, Chicago, IL 60637, USA\\
$^4$Kavli Institute for Cosmological Physics and Enrico Fermi
  Institute, University of Chicago, Chicago, IL 60637, USA\\
$^5$Department of Physics, Yale University, New Haven, CT 06520,
  U.S.A.\\
${^6}$Yale Center for Astronomy and Astrophysics, Yale University, New
Haven, CT 06520, U.S.A.
  }

\begin{document}

\date{Accepted .... Received ...}

\pagerange{\pageref{firstpage}--\pageref{lastpage}} \pubyear{2012}

\maketitle

\label{firstpage}

\begin{abstract}
  We present a new method to identify and characterize the
   structure of the intracluster medium (ICM) in simulated galaxy
   clusters. The method uses the median of gas properties, such as
   density and pressure, which we show to be very robust to the
   presence of gas inhomogeneities. In particular, we show that the
   radial profiles of median gas properties in cosmological
   simulations of clusters are smooth and do not exhibit fluctuations
   at locations of massive clumps in contrast to mean and mode properties. Analysis of simulations shows that distribution of gas properties in a given radial shell can be well described by a log-normal PDF and a tail. The former corresponds to a nearly hydrostatic
“bulk” component, accounting for $\sim$ 99 per cent of the volume, while the tail corresponds to high density inhomogeneities. The clumps can thus be easily identified with the volume elements corresponding to the tail of the distribution. We show that this results in a simple and robust separation of the diffuse and clumpy components of the ICM. The full width half maximum
of the density distribution in simulated clusters is a growing
function of radius and varies from $\sim$ 0.15 dex in cluster centre to
$\sim$ 0.5 dex at $2\,r_{500}$ in relaxed clusters. The small scatter in the width between relaxed clusters suggests that the degree of
inhomogeneity is a robust characteristic of the ICM. It broadly agrees with the amplitude of
density perturbations found in the Coma cluster core. We discuss the origin of ICM density variations in spherical shells and show that less than
20 per cent of the width can be attributed to the triaxiality of the cluster
gravitational potential.
As a link to X-ray observations of real clusters we evaluated the ICM
clumping factor, weighted with the temperature dependent X-ray emissivity, with and without high density inhomogeneities. We argue that these
two cases represent upper and lower limits on the departure of the observed
X-ray emissivity from the median value. We find that the typical value of the clumping
factor in the bulk component of relaxed clusters varies from $\sim$ $1.1-1.2$ at
$r_{500}$ up to $\sim$ $1.3-1.4$ at $r_{200}$ , in broad agreement with recent observations.
\end{abstract}

\section{Introduction}

\begin{table}
 \centering
 \begin{minipage}{60mm}
  \caption{Properties of simulated clusters in our sample at $z=0$.}
  \begin{tabular}{@{}rcc@{}}
  \hline
 Cluster ID & r$_{500c}$ ($h^{-1}$Mpc) & Relaxed (R) \\
            &&Unrelaxed (U)\\  
            &CSF / NR&  CSF / NR\\      
\hline
CL101 &   1.16 / 1.14  & U / U  \\
CL102 & 0.98 / 0.95&U / U\\
CL103 &   0.99 / 0.99  & U / U \\
CL104 &   0.97 / 0.97  & R / R \\
CL105 &   0.94 / 0.92  & U / U \\
CL106 &   0.84 / 0.84  & U / U \\
CL107 &   0.76 / 0.78  & U / U \\
CL3  &   0.71 / 0.70  & R / R \\
CL5        &    0.61 / 0.61 &  R / U \\
CL6        &    0.66 / 0.61  &  U / R\\
CL7        &    0.62 / 0.60  &  R / R\\
CL9        &    0.52 / 0.51  &  U / U\\
CL10       &    0.49 / 0.47  &  R / R\\
CL11       &    0.54 / 0.44  &  U / R\\
CL14       &    0.51 / 0.48  &  R / R\\
CL24       &    0.39 / 0.39  &  U / U\\
\hline
\label{tab:sample}
\end{tabular}
\end{minipage}
\end{table} 

 Hot intracluster gas constitutes $\sim 10 $ per cent of the total
  gravitating mass of galaxy clusters and is the dominant baryonic
  component.  If the gravitational potential of a cluster is static
  then the gas would eventually settle into a hydrostatic
  equilibrium (HSE) with the density and temperature isosurfaces aligned
  with the equipotential surfaces. If in addition the potential is
  spherically symmetric, then all gas thermodynamic properties
  (e.g. density and pressure) are functions of radius only, i.e. the
  intracluster medium (ICM) is homogeneous within a narrow radial
  shell. In reality, both X-ray observations and hydrodynamical
  simulations of galaxy clusters show that the gas is continuously
  perturbed as a cluster forms and the ICM is not perfectly homogeneous
  \citep[see e.g.][]{Mat99,Nag11,Chu12}. Among plausible sources of
  the ICM inhomogeneities are: non-sphericity of the gravitational
  potential; fluctuations of the potential, e.g. due to moving
  subhalos associated with galaxies or subgroups; low entropy gas
  lumps; presence of bubbles of relativistic plasma; turbulent gas
  motions and associated gas displacement; sound waves and shocks,
  etc.

The properties of hot gas in clusters are used to determine the total
gravitating mass of clusters, which is very important for constraining
 cosmological parameters \citep[see
  e.g.][]{Whi93,Hai01,All04,Vik09}.  This is usually done by assuming
that the gas is in HSE and deriving mass profiles
from the temperature and gas density profiles or by using calibrated
mass proxies, such as $Y_X$ parameter \citep{Kra06}. The accuracy of
both approaches is affected by gas inhomogeneities. Therefore,
it is crucial to understand the physical origin of the inhomogeneities
and to find a robust and unambiguous way to characterize and exclude
them from the bulk of the gas.

The level of the ICM inhomogeneity may also depend on the
microphysics, in particular, on the thermal conductivity and viscosity
of the gas and on the topology and magnitude of the magnetic
field. Therefore, the quantitative characterization of the ICM
inhomogeneities could potentially serve as a proxy to these physical
processes.

In this theoretical study we propose a more detailed\footnote{Compared
  to the standard mean radial profiles.}  characterization of the ICM
in numerical simulations. The sample of simulated clusters is
described in Section \ref{sec:simulations}. In Section \ref{sec:bulk}
we introduce a median radial profiles of the gas thermodynamic
properties. The median radial profiles are robust to local
fluctuations and recover the overall smooth radial trends that go
through the peaks of the gas density and temperature distributions in
radial shells. We then introduce an effective measure of the width of
the density distribution (\S\ref{sec:width}) and 
 split the ICM (\S\ref{sec:exclusion}) into a nearly hydrostatic
``bulk'' component, accounting for $\sim$ 99 per cent of the volume, and
non-hydrostatic high density inhomogeneities. The typical gas
velocities in both components and the clumping factor of the ICM are
discussed in Sections \ref{sec:motions} and \ref{sec:cl_factor} respectively. 
The origin of the bulk component inhomogeneities is discussed in Section
\ref{sec:origin}. The sensitivity of the results to the physics
included in simulations is briefly discussed in Section \ref{sec:sens}.
We summarize our results in Section \ref{sec:sum}.

\begin{figure*}
\includegraphics[trim=1.9cm 8.9cm 0mm 8.7cm]{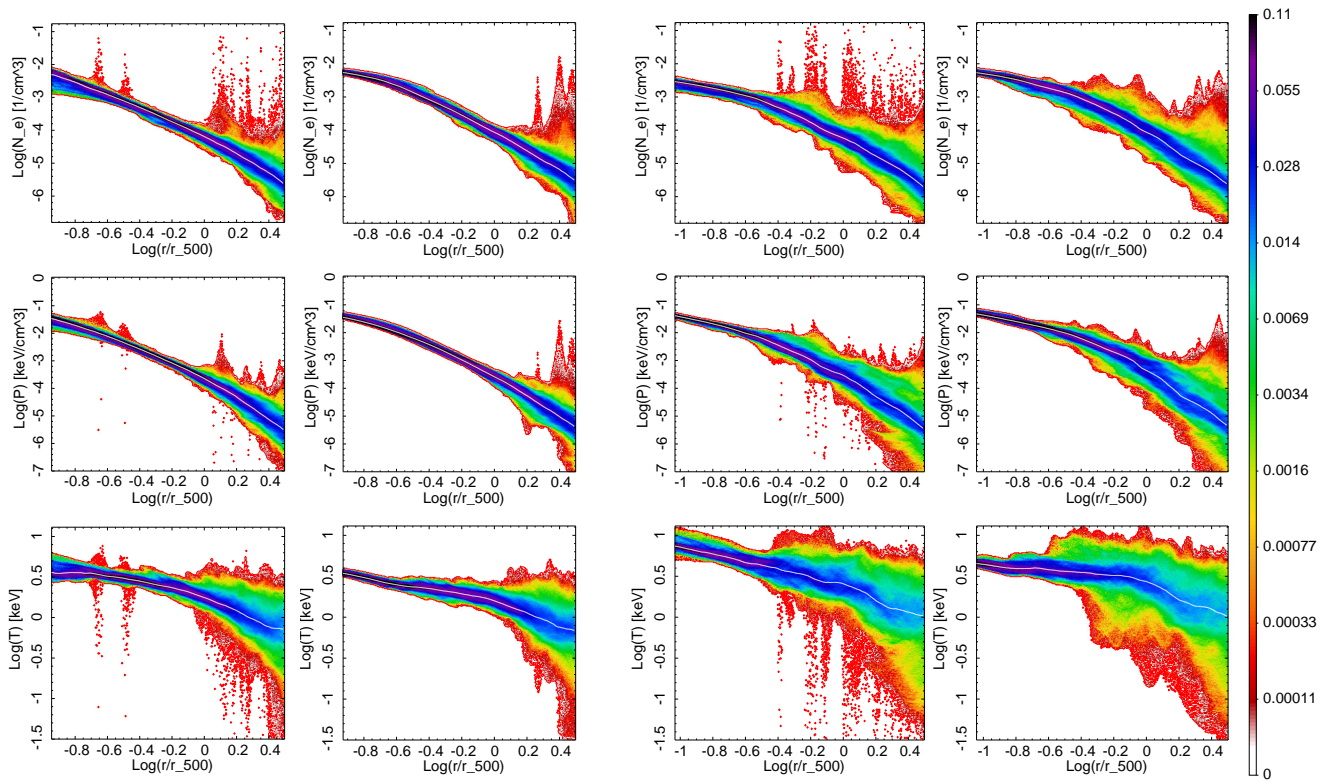}
\caption{Probability Density Function (PDF) of the ICM density,
  pressure and temperature in radial shells as a function of distance
  from the cluster centre. Colour coding: black colour corresponds to
  the highest probability, red colour - to the lowest probability. 
  PDFs are calculated for the relaxed galaxy
  cluster CL7 in CFS (1st column) and NR (2nd column) runs and for
  unrelaxed galaxy cluster CL107 in CSF (3rd column) and NR (4th
  column) runs. The integral of the PDF in each radial shell is
    equal to unity.
  White curves plotted on the top of the PDF show the median
  values of density, pressure and temperature respectively (see
  Section \ref{sec:median}). Strong deviations from the overall smooth
  trend are associated with high density inhomogeneities in the
  ICM. The width of distributions is substantial and increases with
  radius. Note that the width of distributions is larger in unrelaxed
  clusters.
\label{fig:cl7_distr}
}
\end{figure*}

\begin{figure*}
\begin{minipage}{0.49\textwidth}
\includegraphics[trim=0cm 0cm 0mm 3.5cm,width=1\textwidth,clip=t,angle=0.]{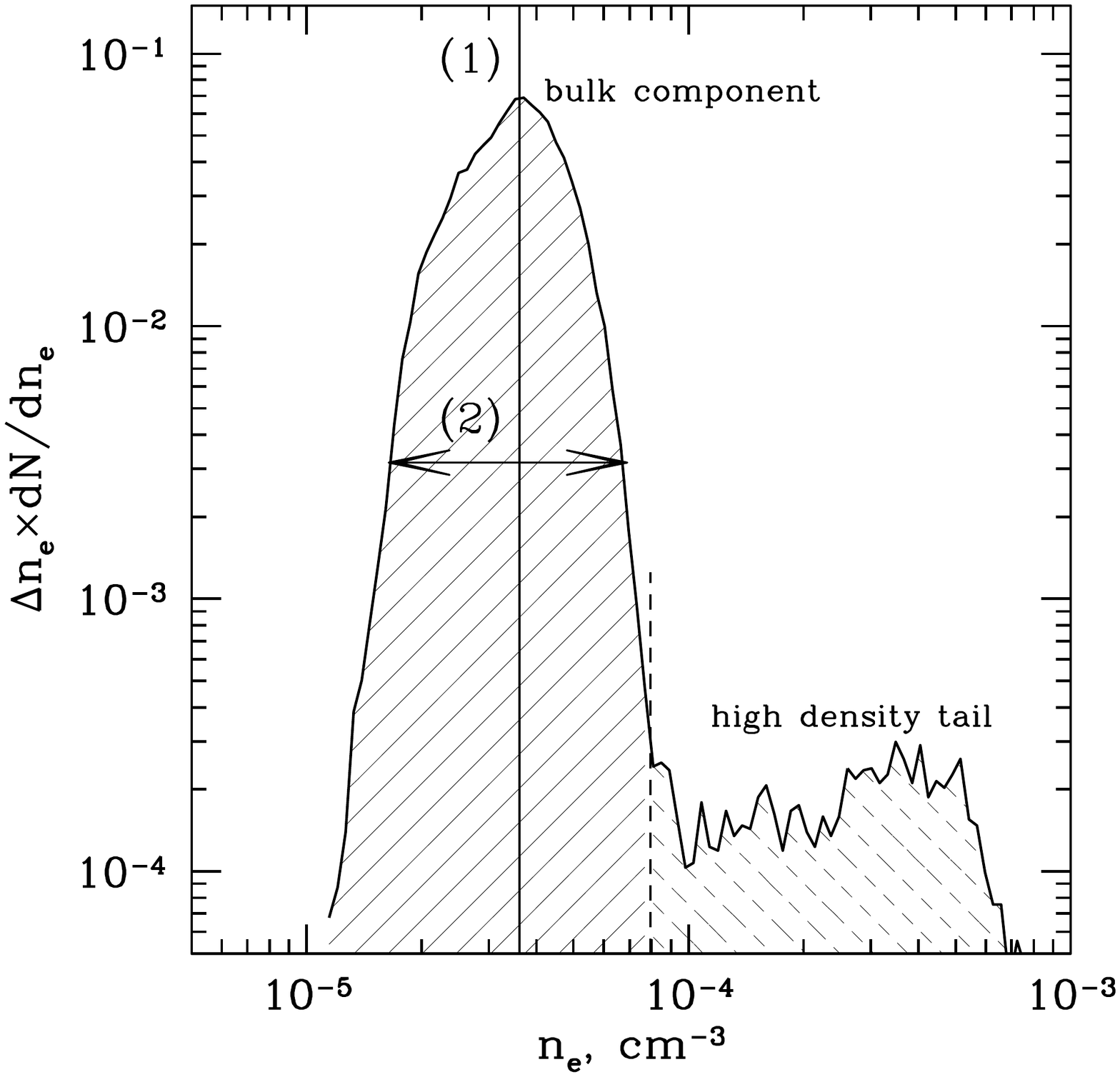}
\end{minipage}
\begin{minipage}{0.49\textwidth}
\includegraphics[trim=0cm 0cm 0mm 3.5cm,width=1\textwidth,clip=t,angle=0.]{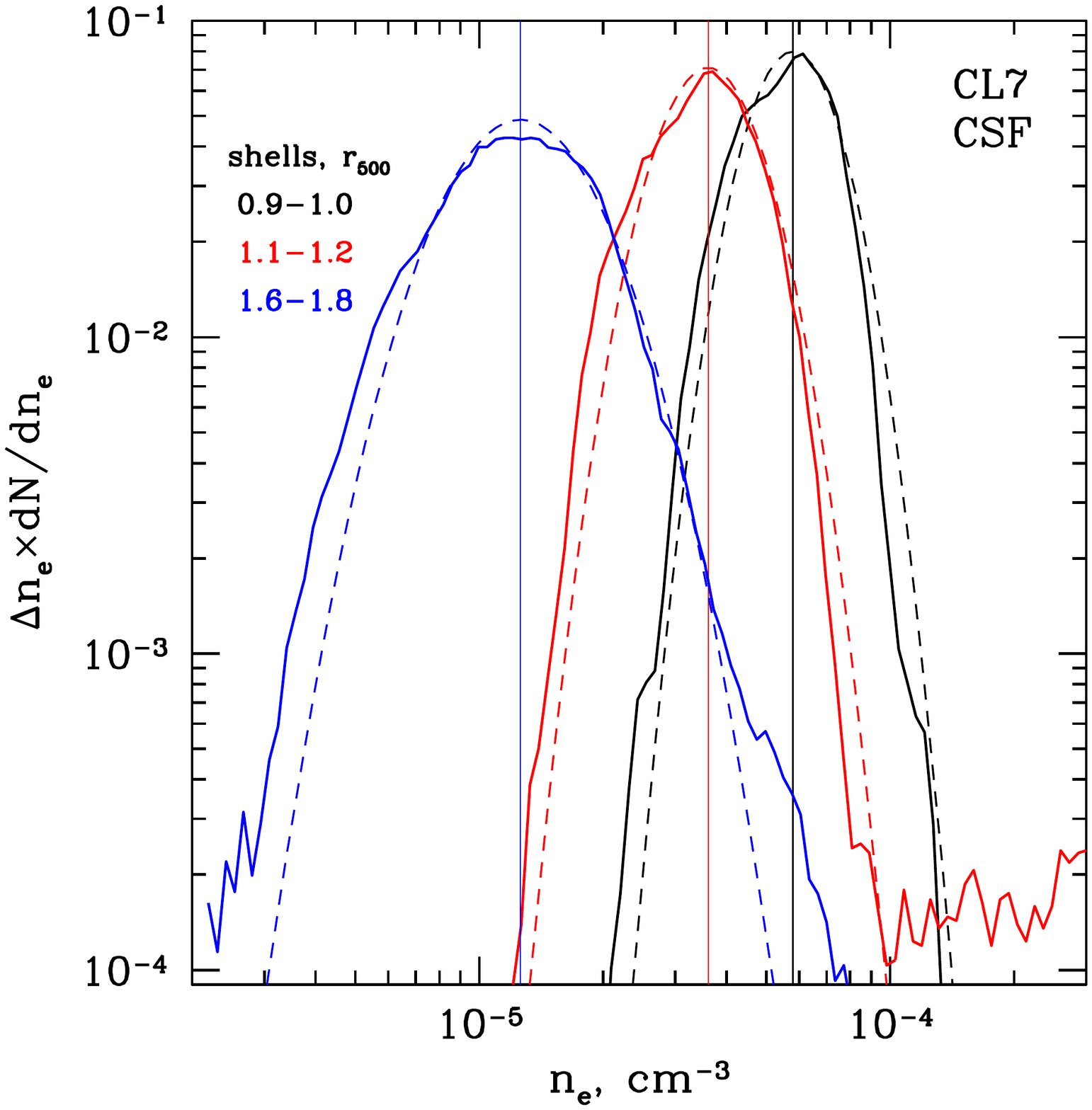}
\end{minipage}
\vspace{-2cm}
\caption{ {\bf Left:} Sketch of ICM description used in the
    paper. The PDF of the density in a radial shell at 1.1-1.2
    $r_{500}$ in the relaxed cluster CL7 (CSF run) is shown with the solid curve. The solid vertical line shows
    the median value of the density (see \S\ref{sec:median}). The ICM
    is divided (see \S\ref{sec:exclusion}) into two components (hatched regions): 
  bulk, volume-filling component and high density inhomogeneities,
  occupying small fraction of the shell volume. The bulk component in
  the paper is characterized by two main parameters: (1) the median
  value of the density and (2) by the width of the density
  distribution. The separation of the components is based on the width
  of the bulk component and on the deviation of the density from the
  median value (see \S\ref{sec:exclusion}). {\bf Right:} Log-Normal 
  approximation of the density PDF. The solid curves show the density
  PDF in three radial shells: 
  0.9-1$r_{500}$, 1.1-1.2 $r_{500}$ (same as in the left panel) and
  1.6-1.8 $r_{500}$. For comparison the dashed
  curves show the log-normal distribution centered at the median
  density value. The Full Width Half Maximum of the log-normal
  distribution is calculated as $\displaystyle W_{10}(n_e)=\log_{10}
  \frac{n_{e,2}}{n_{e,1}}$, where the interval from $n_{e,1}$ to 
  $n_{e,2}$ corresponds to 76 per cent of the shell volume (see
  \S\ref{sec:width}). With these definitions a log-normal distribution
  provides good approximation of the bulk component PDF in each radial shell.
\label{fig:sketch}
}
\end{figure*}

\begin{figure*}
\includegraphics[trim=1.7cm 8.4cm 0cm 8.2cm]{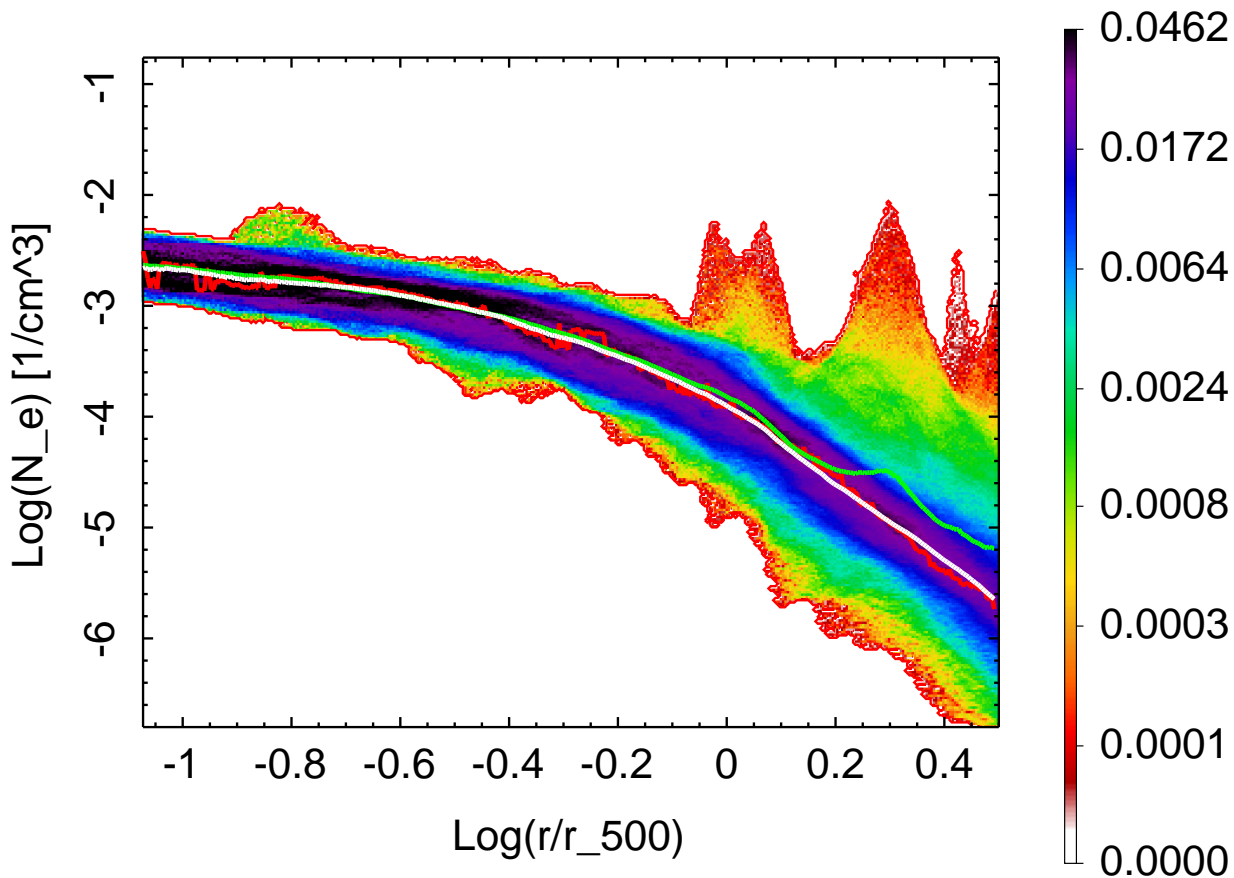}
\caption{PDF (colour coded) of the ICM density in
  radial shells as a function of distance from the centre of the
  CL106 cluster (NR run). Superposed white, green and
  red curves show the median, mean and mode values of density
  respectively. The mode, median and mean radial profiles of density are calculated
  on logarithmic grid  over $r$ with $r_{i+1}=r_if$, where increment f=1.01.
  The {\it median} closely follows the peak of the PDF,
   is a smooth function of radius and does
  not show any wiggles at the radii, where prominent high density
  inhomogeneities are present. The {\it mean} density is reasonably 
  smooth, but it is strongly affected by clumps, which drive it well
  above the PDF peak. The
  {\it mode} value on the other hand by definition coincides with the
  peak of the PDF, but is not smooth.  Its fluctuations reflect
  small variations of the PDF near the maximum. Clearly the median value is an optimal choice if
  one thinks of using it for the HSE
  equation. Indeed, on physical grounds one can claim that only the bulk
   component has a chance to be in the HSE. From the numerical perspective the median is insensitive to the presence
  of inhomogeneities and is smooth, simplifying the calculation of the
  derivatives, needed for the HSE equation.
\label{fig:lcl3_mean_med}
}
\end{figure*}

\section{Simulations and sample of galaxy clusters}
 \label{sec:simulations}
We use a sample of 16 simulated clusters of galaxies at $z=0$
\citep{Nag07a,Nag07b}. The simulations were done using the Adaptive
Refinement Tree N-body+gas-dynamics code \citep{Kra97,Kra02}.
Parameters of a flat $\Lambda$CDM model are $\Omega_m=0.3,
\Omega_b=0.04286, h=0.7$ and $\sigma_8=0.9$. We use two sets of
simulations with the same initial conditions but with different
physics involved in simulations: non-radiative (NR) run without any
radiative cooling or star formation and cooling+star formation (CSF)
run, which includes metallicity-dependent radiative cooling, star
formation, supernova feedback and UV background. These 16
  clusters with virial masses ranging from $\sim7\times 10^{13}$ to
  $2\times 10^{15}~h^{-1}~M_\odot$ were selected from low-resolution
  simulations and resimulated at higher resolution. The initial
  selection was not aimed to balance between relaxed and unrelaxed
  clusters. The division of the sample into relaxed and unrelaxed
  subsamples (see Table \ref{tab:sample}) was done in \citet{Nag07b}
  by visually examining the morphology of mock X-ray images.

Instead of using full AMR mesh hierarchy, for convenience we sample the
hydrodynamical data using $4\times 10^7$ random data points within the
sphere of radius of 5 Mpc/h centered on the cluster centre, defined as
the position of the most bound particle in the simulation
box. The simulated volume is sampled with a weight $\propto 1/r^2$,
where $r$ is the distance from the centre. This sampling is
uniform in azimuthal and polar directions and provides equal
number of points per spherical shell of a given thickness. As the
result the 3D density of sampling points is highest at the centre. 

 Using these data we generated Probability
  Density Function (PDF) of the ICM thermodynamic quantities in
  a set of radial shells for all clusters in the CSF and NR runs. The examples of the PDF for the
  relaxed cluster CL7 and unrelaxed cluster CL107 are shown in
  Fig.~\ref{fig:cl7_distr}. The colour changing from red to dark-blue characterises
  the increasing volume-weighted probability of finding gas with a given density
  (or pressure/temperature) at a given radius. 
Overall radial trends of all thermodynamic properties are apparent
from these plots. Moderate amplitude fluctuations of the ICM
properties around these trends are visible as blue/green
bands, which become broader with radius. Typically these bands account
for $\sim$ 99 per cent of a shell volume. Below we refer to these bands as
a volume-filling bulk component of the gas. Finally, high density
inhomogeneities, occupying very small fraction of volume, are seen as
red spikes. Another representation of both components is shown in the
left panel in Fig. \ref{fig:sketch}.

\section{Characterizing the bulk component of the
  ICM}

\label{sec:bulk}
\label{sec:median_width}
\subsection{Median profiles}
\label{sec:median}

\begin{figure*}
\begin{minipage}{0.49\textwidth}
\includegraphics[trim= 0mm 0cm 0mm 3.5cm, width=1\textwidth,clip=t,angle=0.]{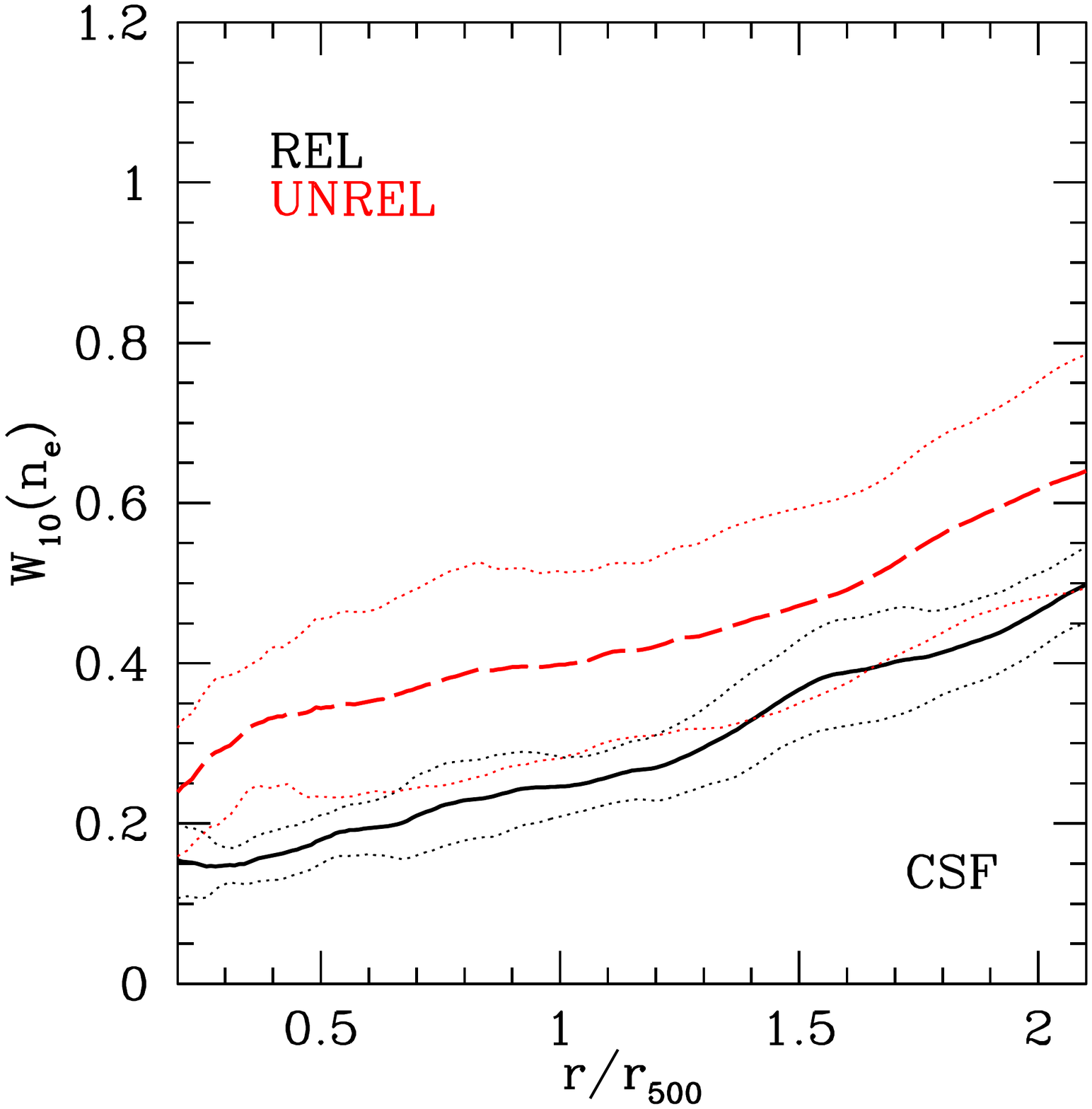}
\end{minipage}
\begin{minipage}{0.49\textwidth}
\includegraphics[trim= 0mm 0cm 0mm 3.5cm,width=1\textwidth,clip=t,angle=0.]{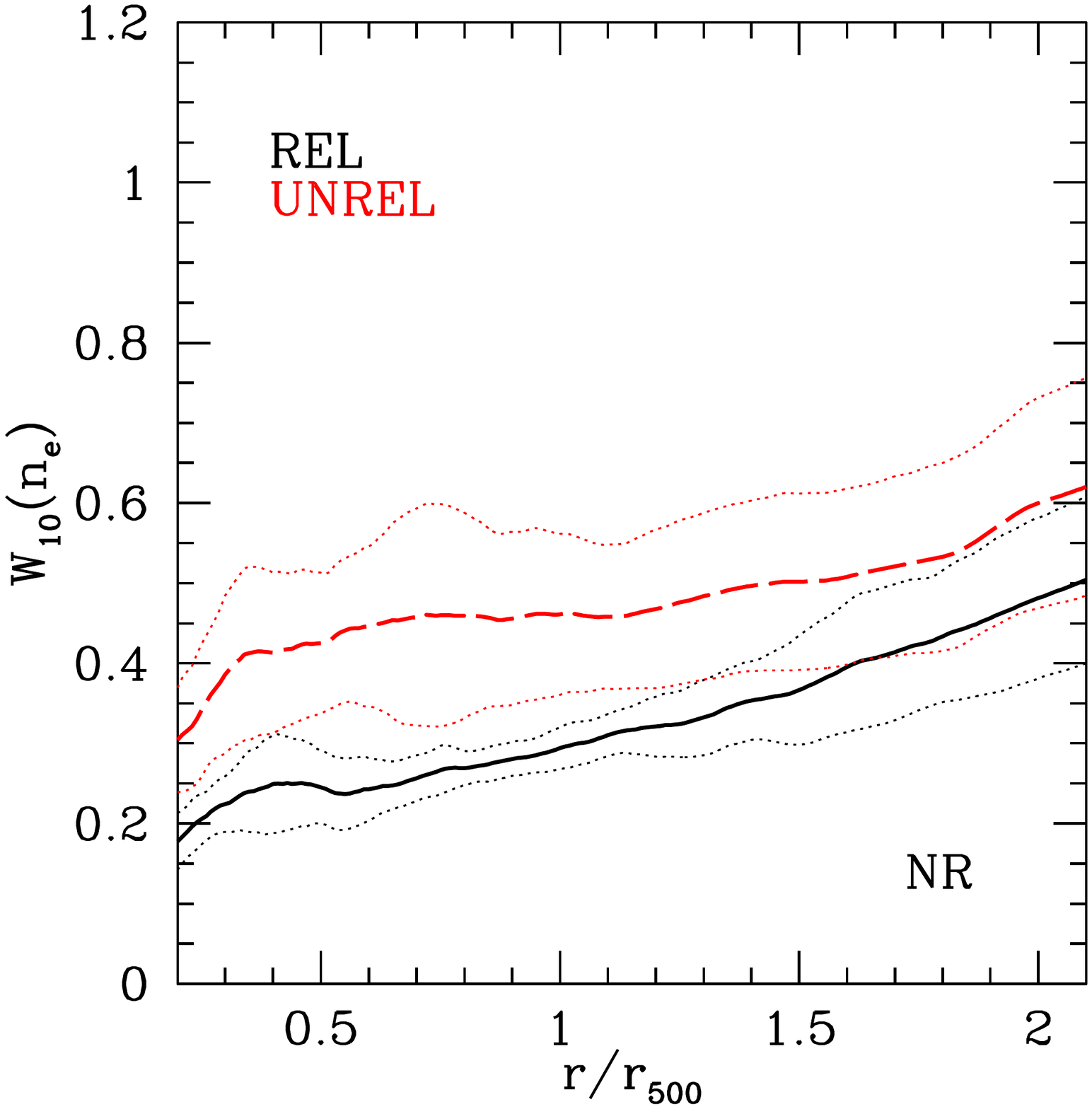}
\end{minipage}
\begin{minipage}{0.49\textwidth}
\includegraphics[trim= 0mm 5cm 0mm 7cm,width=1\textwidth,clip=t,angle=0.]{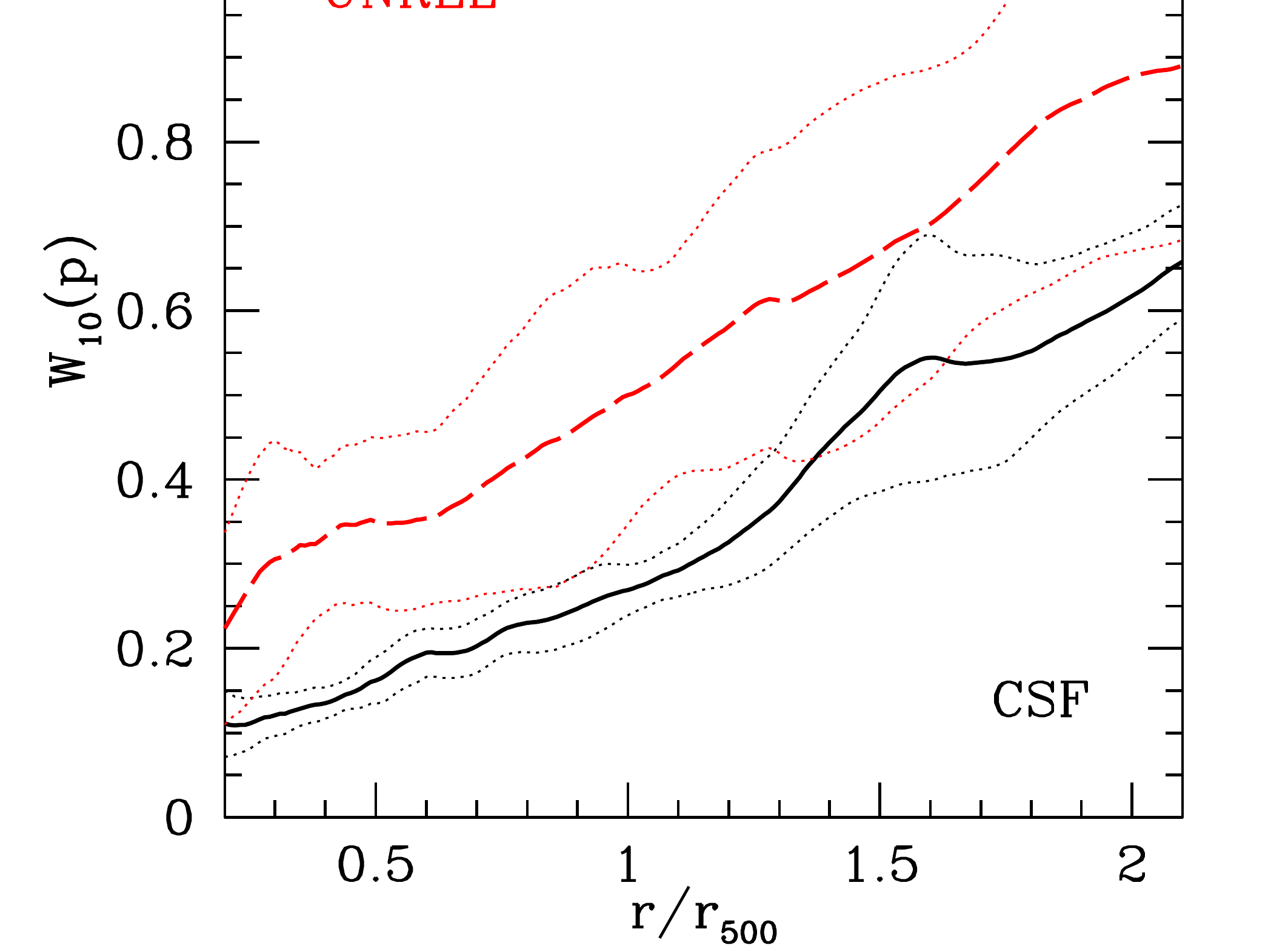}
\end{minipage}
\begin{minipage}{0.49\textwidth}
\includegraphics[trim= 0mm 5cm 0mm 7cm,width=1\textwidth,clip=t,angle=0.]{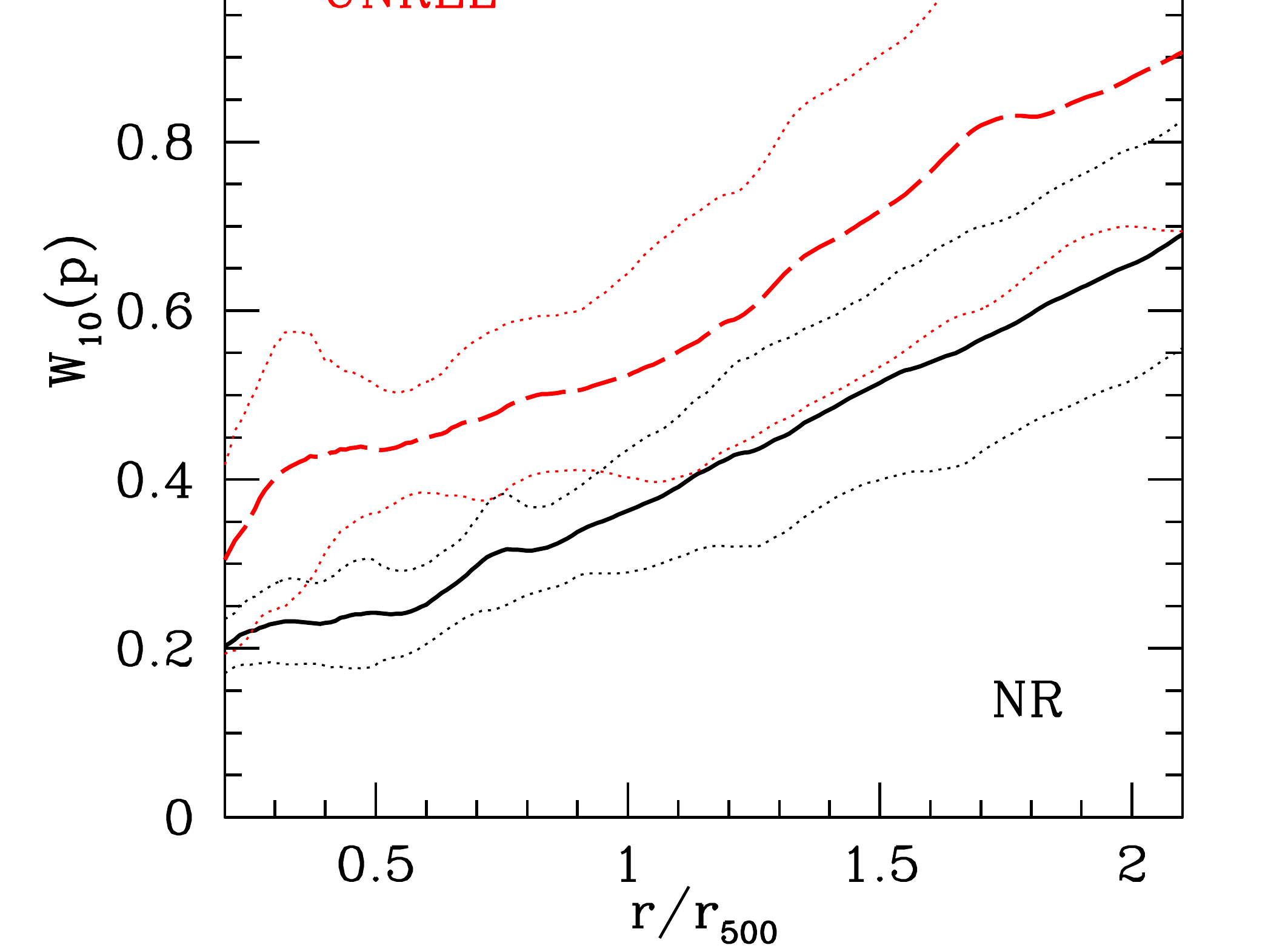}
\end{minipage}
\caption{Width $W_{10}$ of the density (top panels) and pressure
  (bottom panels) distributions in a radial shell as a function of
  distance from the cluster center for CSF (left column) and NR (right column)
  simulations. The width is averaged over the sample of relaxed (black
  solid curves) and unrelaxed (red dashed curves) clusters. Thick
  solid and dashed curves correspond to the sample-averaged value and thin dotted
  curves show the scatter from cluster to cluster.
\label{fig:width}
}
\end{figure*}

First, we would like to characterise overall radial profiles of the ICM
thermodynamic properties, representing the bulk volume-filling
component of the ICM. A primary application of these profiles is the
cluster mass measurements via HSE equation. 

Usually hot gas is characterized by the mean radial profiles of
density and pressure obtained under the assumption of spherical
symmetry. These profiles are used to determine the mass of the
cluster using the HSE equation \be
\frac{1}{\rho}\frac{dP}{dr}=-\frac{GM}{r^2},
\label{eq:he}
\ee where $\rho$ and $P$ are radial profiles of gas density and pressure
respectively and $M$ is the total gravitating mass of the
cluster. If the pressure is due to the thermal gas pressure, then
$P=nkT$ and $\rho=\mu m_p n$, where $\mu$ is the mean atomic weight of
the gas particles, $m_p$ is the proton mass, $k$ is the Boltzmann constant and $n$ is the total
particles density. Throughout the paper we use $\mu=0.588$. This procedure requires differentiation of the pressure over
the radius. Thus the issue of a robust way of calculating radial pressure
profiles is especially important. Various high density
  inhomogeneities affect the measurements of the mean radial profiles. Moreover, while the bulk of
the gas may be close to the hydrostatic equilibrium in the cluster
potential, the high density inhomogeneities are obviously far from 
equilibrium. Therefore in order to avoid spurious variations of the
mean profiles due to high density inhomogeneities one has to excise them from
the data. 
Often, when analyzing simulated data, the high density gas
clumps are removed by introducing some threshold
values in the density/temperature values and excising the regions
where the ICM parameters violate these thresholds \citep[e.g.][]{Lau09,Vaz11,Fab11}. The radial profiles
are then calculated by averaging the density (or pressure/temperature),
over the remaining volume. However, the resulting mean profiles are
sensitive to the particular procedure of clump removal. High density inhomogeneities can significantly shift the
mean density or temperature, causing distortions in the mean pressure.
We instead are seeking a method which will be robust with respect
to the presence of inhomogeneities and does not require fine tuning of the clump removal procedure.

We propose to use median radial profiles of density, temperature and
pressure instead of their mean quantities as is most commonly
done. Given N particles in a radial shell the calculation of the
median is reduced to sorting particles in ascending/descending order
and taking the value corresponding to a particle with index
N/2.\footnote{In our case all particles are uniformly distributed over
  the volume and median is calculated with unit weight, automatically
  giving us volume-weighted median. In case of SPH simulations one
  should use weights inversely proportional to local density to obtain
  volume-weighted median instead of the mass-weighted median, since
  particles are distributed non-uniformly: the denser the region is
  the more particles it contains.} White curves in
Figs. \ref{fig:cl7_distr} and Fig.~\ref{fig:lcl3_mean_med} show
resulting median radial profiles.  These median profiles can be
favorably compared (Fig.~\ref{fig:lcl3_mean_med}) to the mean and mode
profiles. The median profile is smooth and follows well the peak of the
PDF even when contamination by high density gas inhomogeneities is
very severe.  Of course, this is true only as long as the
  fraction of volume occupied by the high density component is small.
The mean density profile is reasonably smooth, but it is strongly
affected by clumps, which drive it well above the PDF peak. The 
  mode value by definition coincides with the peak of the PDF, but it
  is not smooth.  Its fluctuations reflect (possibly small) variations
  of the PDF near the maximum.  

Clearly the median value is an optimal choice if one thinks of using it
for the hydrostatic equilibrium equation. It can be calculated
straightforwardly from the PDFs in spherical shells without need to
select or tune procedure of high density clumps removal. It
characterizes directly the properties of the bulk component of the ICM
and is not affected by the presence of high density inhomogeneities,
as long as their volume fraction is small. The median pressure profile
is a smooth function of radius, making it very suitable for the
HSE equation.

\subsection{Width of density and pressure distributions}
\label{sec:width}
We now proceed with the evaluation of the width of the bulk component
distribution. Fig. \ref{fig:sketch} (the right panel) shows the ICM density PDF in several radial
  shells. In some shells the density distributions  are very asymmetric due to the presence of high density
tails. However, if we exclude tails, the remaining distribution can be
reasonably well described by a log-normal distribution around the
median value
 \be
P(\ln x)d \ln x=\frac{1}{\sqrt{2\pi}\sigma} e^{\disp-\frac{(\ln x-\ln
    x_0)^2}{2\sigma^2}}d \ln x,
\label{eq:logn}
\ee where $x_0$ is the median value, as illustrated in
Fig. \ref{fig:sketch} \citep[see also][]{Kaw07}. Even if we exclude high density tails, the
  distributions are quite broad, especially at large $r$. In Section
  \ref{sec:origin} below we argue that the contribution of the overall
  ellipticity of the potential to the calculated
  width of the distribution does not exceed 20 per cent in relaxed clusters. We therefore refer
  to the broadening of the ICM density distribution in a radial shell
  around the medial values as ``perturbations''.

\begin{figure}
\includegraphics[trim=0.8cm 5.5cm 0cm 3.3cm,width=0.5\textwidth,clip=t,angle=0.]{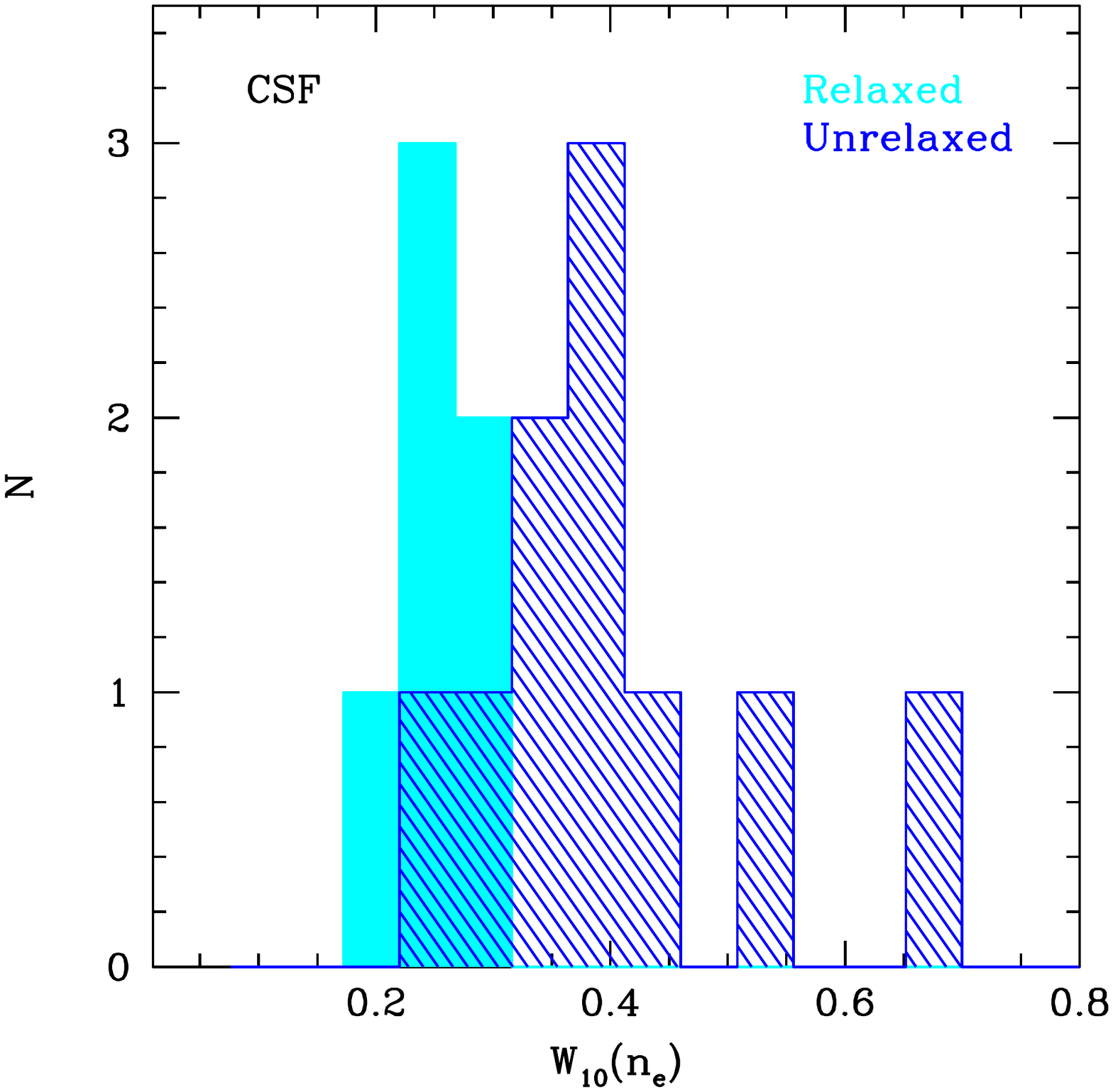}
\caption{Histogram of the width of the gas density
  distribution $W_{10}(n_e)$  at $r_{500}$ for the CSF sample of relaxed clusters
  (cyan solid) and unrelaxed clusters (blue dashed). One can see
    two peaks in the distribution. The significance of the difference
    in width between two samples is $\sim 2.5 \sigma$. The width $W_{10} \sim
    0.3$ at $r_{500}$ divide clusters into relaxed/unrelaxed
    samples. This way of classification is independent of the
    projection effects.
\label{fig:width_distr}
}
\end{figure}

Let us calculate the width of the density and pressure distributions -
another important characteristic of the bulk component. We are seeking
the procedure which is not very sensitive to the presence or absence
of high density tail in the distribution. While the
shape of the PDF of the bulk component is close to log-normal, small
deviations are present. We therefore introduced an ``effective full
width half maximum'', $W_{10}$, as a proxy to the distribution
width. The value of $W_{10}$ is calculated as follows.  For each
radial shell we find the value of density $n_{e,1}$ such that 12 per cent of
points (volume) in this shell have the density smaller than
$n_{e,1}$. Similarly we find $n_{e,2}$ such that 12 per cent of points
(volume) have higher density than $n_{e,2}$. $W_{10}$ is then
defined as:
\be
W_{10}(n_e)=\log_{10} \frac{n_{e,2}}{n_{e,1}}.
\ee
$W_{10}$ characterizes the logarithmic interval (10-based), which
contain 76 per cent of points. Clearly, for a pure log-normal distribution
$W_{10}$ is equal to FWHM ($\log_{10}$-based). This definition is also
convenient, since numerically $\displaystyle
W_{10}=\frac{2\sqrt{2\ln2}}{\ln 10}\sigma\approx 1.02\sigma$, where
$\sigma$ is the standard deviation (natural log based) of the
log-normal distribution. With this definition of the width, the
log-normal distribution, centered at the median value, provides
reasonably good approximation of the bulk component PDF (see Fig. \ref{fig:sketch}).

Fig. \ref{fig:width} shows the width of the distributions
$W_{10}(n_e)$ and $W_{10}(P)$ averaged over a sample of relaxed and
unrelaxed galaxy clusters in CSF and NR runs. First we notice a strong
increase of the width with radius $r$, which indicates that the gas
is more inhomogeneous towards cluster outskirts. Scatter in the width
from cluster to cluster is relatively small (especially for relaxed
clusters). Pressure distributions are broader than density
distributions at $r>r_{500}$, while at $r<r_{500}$ their widths are
similar. Strong growth of the pressure width can be an indication of
the gas deceleration at these radii. We refer readers to Section \ref{sec:origin} for
discussion on possible physical origin of the width of density and
pressure distributions.

One can notice the tendency of unrelaxed clusters to have broader
distribution of density and pressure than the relaxed clusters. This
is not surprising, since any strong merger should perturb the density
distribution. This
tendency is even more clear if one looks at the histogram of
$W_{10}(n_e)$ at certain distance from the
centre. Fig. \ref{fig:width_distr} shows the corresponding histogram 
 calculated at $r_{500}$ for relaxed
and unrelaxed clusters in CSF run. Even though samples
are small, the significance of the difference in width between two samples is
$\sim 2.5 \sigma$. It suggests a possibility of using the width
$W_{10}$ at certain radius (e.g. $r_{500}$) as a criterion for an
automatic division of clusters into broad relaxed/unrelaxed
groups. Note that this way of classification is independent of the
projection effects. Classification of all clusters into relaxed or
unrelaxed objects in \citet{Nag07a} is based on the visual inspection of
the mock X-ray images (see their Section 2). Further inspection of CL6
and CL9 clusters, identified as unrelaxed in \citet{Nag07a}, but
having relatively narrow density distribution ($W_{10}\le 0.3$) shows
that these objects can equally well be attributed to a class of relaxed objects within
  $r_{500}$.

\section{Method to select high density inhomogeneities}
\label{sec:exclusion}

\begin{figure}
\includegraphics[trim=1.7cm 5.5cm 0cm 5cm,width=0.52\textwidth,clip=t,angle=0.]{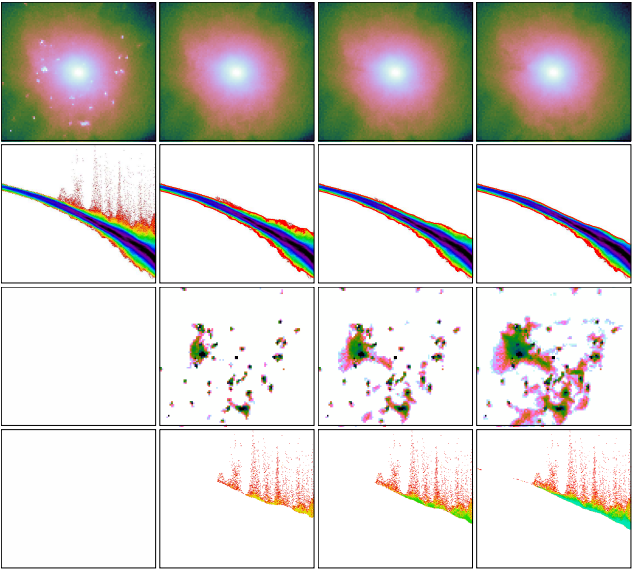}
\caption{Splitting of the ICM into high density inhomogeneities and bulk
  component of gas assuming different criteria of
  separation. Particles with $\displaystyle \log_{10} n_e >
\log_{10} \{n_e \}+ f_{cut}\sigma_{10}$, where $\sigma_{10}$ is the
width of density
  distribution at each radial bin, are assigned to the high density
  component (Section \ref{sec:exclusion}). Remaining
  particles are assigned to the bulk component.  Columns from the left
  to the right: $f_{cut}=\infty$(total gas) $, 4.5, 3.5, 2.5$. {\bf Top
  row:} projected density maps of the bulk component of gas, {\bf
  second row:} PDF of the ICM density of the bulk
  component in radial shells as a function of distance from the
  cluster centre, {\bf third row:} density maps of the high density
  inhomogeneities, {\bf bottom row:} density distributions of the high
  density component in radial shells as a function of distance from
  the cluster centre. The figure suggests that $f_{cut}=3.5$ provides an optimum threshold for the division
  into bulk component and high density inhomogeneities --
  it removes only small fraction of particles and at the same time
  does not leave any obvious inhomogeneous features in the PDF.
\label{fig:crit_lcl4}
}
\end{figure}

Once we know the median density profile and the width of density
distribution it is easy to separate the bulk component from the tails
of the distribution. It
additionally allows to study the properties of both components
separately. We propose to use the following
criterion  of separation: particles with $\displaystyle \log_{10} n_e >
\log_{10} \{ n_e \}+ f_{cut}\sigma_{10}$ are assigned to the high density tail,
while all remaining particles belong to the bulk component. Here
$\{ n_e \}$ is the median value of the density and $\displaystyle
\sigma_{10}=\frac{W_{10}}{2\sqrt{2\ln2}}$ is the standard deviation
($\log_{10}$ based) of the density distribution. The choice of
$f_{cut}$ is rather arbitrary. Experiments with
simulated galaxy clusters in our sample show that $f_{cut}=3.5$ works
well for both NR and CSF simulations.
Clearly, by varying $f_{cut}$ one can select only the densest clump cores or the clumps
together with the surrounding elevated density regions. These points
are illustrated in Fig. \ref{fig:crit_lcl4}, where we show density
distributions and projected density maps obtained assuming different
values of $f_{cut}=
\infty$ (initial maps), $4.5, 3.5$ and $2.5$. In the case of
$f_{cut}=2.5$ we select not only ``bona fide" high density tails, but
also partially exclude particles, which belong to log-normal
  density distribution and characterize the bulk of the gas. In the case of $f_{cut}=4.5$ we
separate only the top of densest clumps and attribute some of the
substructure clearly related to clumps to the bulk gas.

\begin{figure}
\includegraphics[trim=0.7cm 5.5cm 0cm 3cm,width=0.51\textwidth,clip=t,angle=0.]{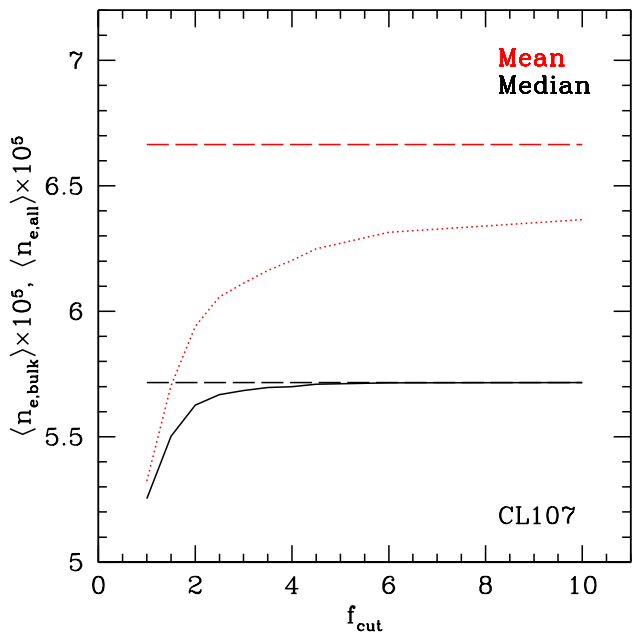}
\caption{Sensitivity of the median (black) and mean (red) values of
  the density to the cutoff value, used to separate the bulk and high
  density components. Dashed curves show the mean and median
    values of the total density from simulations $\langle
    n_{e,all}\rangle$ at $r_{500}$. Dotted
    and solid curves show density of the bulk gas at $r_{500}$ as
    a function of $f_{cut}$. Large
    $f_{cut}$ excludes only the densest inhomogeneities, while very
    small $f_{cut}$ can exclude in addition slightly overdense gas,
    belonging to the bulk component (see Fig. \ref{fig:crit_lcl4}). Closeness of the black
    curves over a broad range of $f_{cut}$ implies that the median is
    not sensitive to the particular value of the cutoff, unless it is
    very low. In contrast, the value of the mean density is much more
    strongly affected by the presence of rare high density
    inhomogeneities and by the choice of the threshold $f_{cut}$ used
    to excise the inhomogeneities.
\label{fig:meanmedcut}
}
\end{figure}

Here it is important to point out that the median profiles of the bulk
component are essentially insensitive to the $f_{cut}$ value in
comparison with the mean profiles. Fig. \ref{fig:meanmedcut}
illustrates this point for one cluster from the sample. We calculate median
and mean radial profiles of the density in shells, removing different
fractions of dense clumps (varying
$f_{cut}$). Fig. \ref{fig:meanmedcut} shows the mean and median values
of the density of gas from simulations $n_{e,all}$ (both bulk gas and
inhomogeneities) and the
density of gas with excluded high density tail $n_{e,bulk}$ at
$r_{500}$ as a function of cutoff $f_{cut}$. We see that once the
$f_{cut} \ge 3$, the median is less sensitive to the presence of
inhomogeneities and to various ways to exclude them than the mean.

\section{Gas motions of the bulk component and the high density 
inhomogeneities}

\label{sec:motions}

\begin{figure*}
\begin{minipage}{0.49\textwidth}
\includegraphics[trim=0cm 0.5cm 0cm 3.5cm,width=1\textwidth,clip=t,angle=0.]{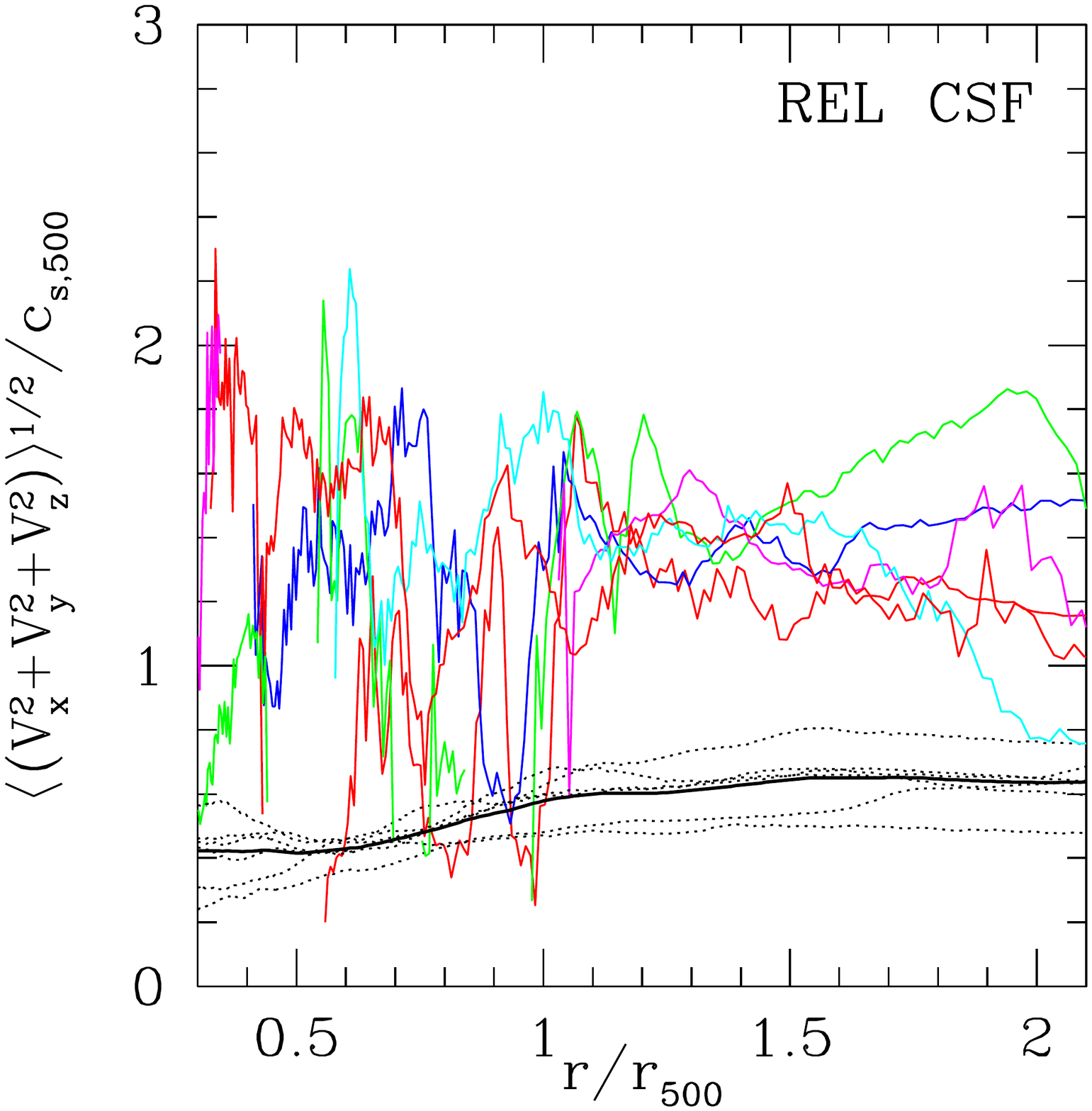}
\end{minipage}
\begin{minipage}{0.49\textwidth}
\includegraphics[trim=0cm 0.5cm 0cm 3.5cm,width=1\textwidth,clip=t,angle=0.]{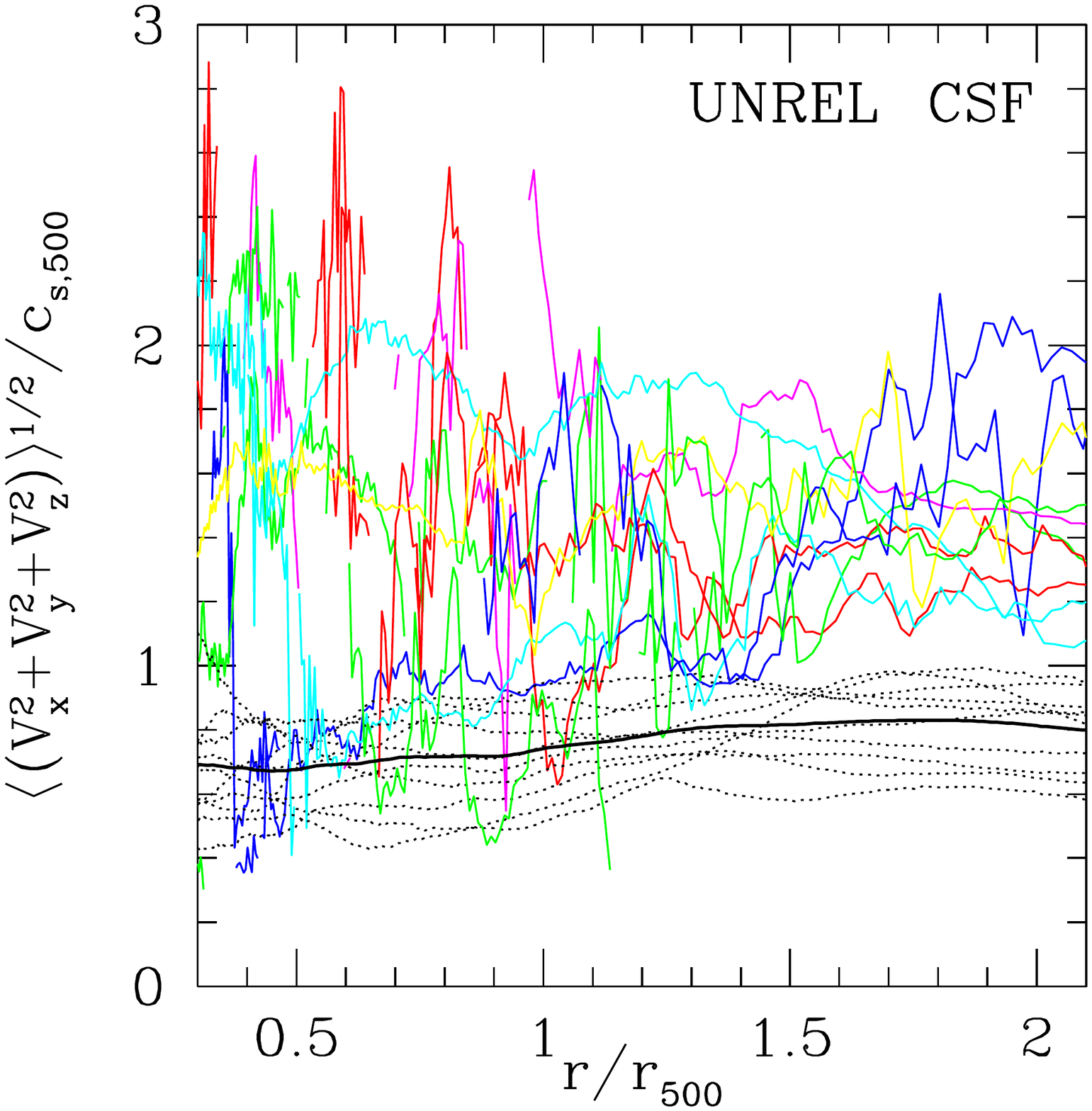}
\end{minipage}
\vspace{-2cm}
\caption{Ratio of the RMS velocity amplitude (averaged in shells) and
  the sound speed at $r_{500}$ as a function of
  distance from the cluster centre. The sound speed at $r_{500}$ is
  calculated as $\disp c_{s,500}=\disp \sqrt{\gamma k T_{500}/ \mu m_p}$ (see
  Section \ref{sec:motions} for details). {\bf Solid coloured curves:}
  ratio for the dense inhomogeneities only, different colours correspond
  to different clusters. {\bf Dotted black curves:} ratio for the bulk
  gas component. {\bf Thick solid black curves:}
  ratio for bulk component of gas, averaged over the sample of relaxed
  (left panel)
  or unrelaxed (right panel) clusters. Note that on average clumps are moving with
   larger velocity than the gas in the bulk component. 
\label{fig:vmean}
}
\end{figure*}

After splitting the ICM into two components it is easy to calculate
characteristic velocities of the bulk component and of the high
density inhomogeneities. As the reference velocity we use velocity averaged over the cluster core. Since in the present simulations the central $\sim$ 300
kpc region is strongly affected by the excessive gas cooling
\citep[e.g.][]{Lau12} that produces unphysical clumps moving
  with very high velocity,
the average gas velocity within a wide radial shell at $400<r<500$ kpc
was subtracted from the velocity field. Our experiments with
  different choices of region used to calculate the reference velocity
  have shown that final results are only weakly sensitive to this choice. The RMS velocity amplitude
was calculated as $\sqrt{\langle V_x^2+V_y^2+V_z^2\rangle}$, where
$\langle\rangle$ denotes averaging over the particles within a
shell. Fig. \ref{fig:vmean} shows the ratio of the RMS velocity and
the sound speed evaluated at $r_{500}$
\be
\disp c_{s,500}=\disp \sqrt{\frac{\gamma k
T_{500}}{\mu m_p}},
\ee
where $T_{500}$ is the
gas temperature at $r_{500}$, $\gamma$ is the adiabatic index (for
ideal monoatomic gas it is 5/3), $k$ is the Boltzmann constant, $\mu$
is the mean atomic weight and $m_p$ is the proton mass. One can notice
that (i) RMS velocity of the bulk component has a very regular
behaviour with distance from the cluster centre, while the velocities
of the high density component vary strongly; (ii) the scatter of the
velocities between individual clusters is small for the bulk
component and is on the contrary very large for the dense inhomogeneities; (iii)
on average dense clumps move faster than the bulk component. Such
behaviour of the RMS velocity in both components is not
surprising. The bulk component is close to the HSE, while high density inhomogeneities are far from
equilibrium. Note that for high density inhomogeneities $\disp\frac{V_{rms}}{c_{s,500}}\approx 1$ at $r_{500}$. This means that the
  clumps' kinetic energy is about the same as the thermal 
  energy per unit mass in the bulk component. This is expected for the
  gas that has been heated by the thermalization of the bulk gas motions
  with velocities comparable to the observed velocities of the
  strongly overdense clumps \citep{Fel66}.

\begin{figure}
\includegraphics[trim=0.5cm 5cm 0cm 3cm,width=0.5\textwidth,clip=t,angle=0.]{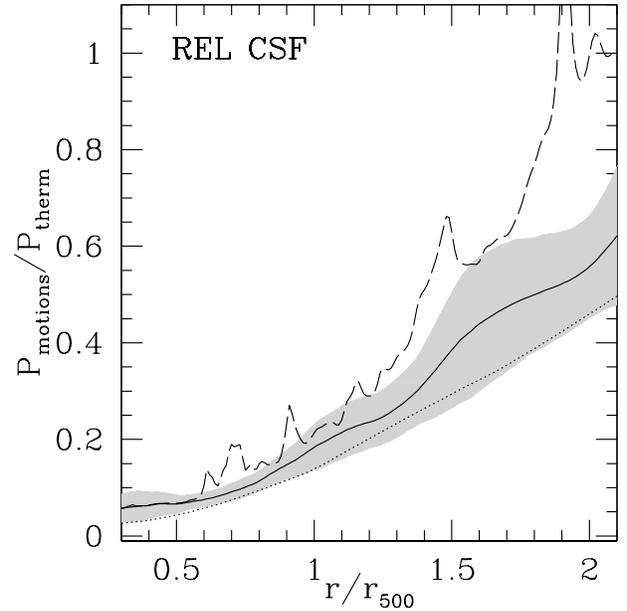}
\caption{ Ratio of the pressure due to gas motions and the gas thermal
  pressure as a function of distance from the cluster centre, averaged
  over a sample of relaxed clusters in the CSF run. {\bf
    Dashed curve:} pressure ratio for total gas distribution (bulk
  plus high density inhomogeneities) from simulations. {\bf Solid curve:} the
  ratio for the bulk component. {\bf Gray shadow:} scatter of the
  ratio for the bulk component. {\bf Dotted curve:} pressure ratio from
  \citet{Lau09}. See Section \ref{sec:motions} for
  details.
\label{fig:pmean}
}
\end{figure}

We also gauge how strongly high density inhomogeneities affect the
ratio of thermal pressure and the ``pressure'' due to stochastic gas
motions. The ratio of pressures is given by
\be
\disp \frac{P_{motions}}{P_{thermal}}=\disp\frac{\langle\disp\frac{1}{3}\rho V_{amp}^2\rangle}{\{nkT\}},
\ee    
where $V_{amp}$ is the RMS velocity amplitude (see
Fig. \ref{fig:vmean}), $\langle \rangle$ and $\{ \}$ denote mean and
median values of pressure respectively. As an example, we show this
ratio as a function of $r$ for relaxed CSF clusters in
Fig. \ref{fig:pmean}. Contribution of the pressure due to gas motions
to the thermal pressure is $\sim$ 5 per cent in the cluster centre and
increases with radius. Exclusion of high density
inhomogeneities leads to a significantly smaller ratio at
$r>r_{500}$. Once again this demonstrates that the bulk
component is much closer to the HSE, than the
ICM inhomogeneities. Comparison with previous results from
\citet{Lau09} shows a broad agreement (dotted and dashed curves in
Fig. \ref{fig:pmean}). Small discrepancies in radial profiles are
mostly due to different ways to subtract the mean velocity from the
total velocity field. We subtract the mean velocity in radial shell
$400<r<500$ kpc as described above, while \citet{Lau09} subtract mean
velocity in each radial shell. Also discrepancies in pressure ratio
are due to different procedures of clump exclusion and slight
distinction between median and mean thermal pressures.

\section{Clumping factor with and without dense inhomogeneities}
\label{sec:cl_factor}
We calculated the clumping factor, which characterizes distortions in the
X-ray flux from a given shell, relative to the flux calculated using
the median temperature and density values in the same shell. This factor is:
\be
C_{X}=\disp\frac{\disp\langle\Lambda(T)n_e^2\rangle}{\disp\{\sqrt{\Lambda(T)}n_{e}\}^2},
\label{eq:clfac}
\ee
where $\Lambda(T)$ is emissivity in a given energy band as a function
of temperature, $\langle \rangle$ and $\{ \}$ denote mean and median values respectively. The emissivity as a function of
temperature was calculated for the 0.5-2 keV energy band - where the
sensitivity of current X-ray observatories is close to maximal for a
typical cluster spectrum.

The choice of the emissivity-weighted clumping factor $C_X$ instead of
the ``classical" clumping $\langle n_e^2 \rangle/\langle n_e
\rangle^2$ is motivated mainly by the two reasons. First, emissivity
factor 
automatically takes care of the exclusion of the densest and coldest
gas clumps present in simulations. Thus we do not need to
introduce a cut over the temperature to calculate clumping for X-ray
emitting gas \citep[e.g.][]{Nag11}. Secondly, median value in
denominator in eq. (\ref{eq:clfac}) is characteristics of the hydrostatic component in the
ICM. Therefore, clumping factor $C_X$ reflects the increase of the
surface brightness due to inhomogeneities in the ICM relative to the
surface brightness in an ideal hydrostatic situation. We conclude that
calculation of the clumping factor using eq. \ref{eq:clfac} has good
physical motivation.

Fig. \ref{fig:clfac} shows clumping factor
calculated for the total gas in simulations and for the bulk nearly
hydrostatic component. As expected, the exclusion of high density inhomogeneities significantly modifies
the clumping factor: clumping factor becomes smaller, especially at
$r>r_{500}$ and smoother with radius.  Both curves determine the upper
and lower limits on the boost of X-ray flux over the flux in the bulk
gas we expect to find in X-ray
observations. \citet{Nag11} calculated the clumping factor for gas
with $T>10^6$ K. Such a temperature cut partially excludes high density
inhomogeneities in the ICM. Comparison shows that the clumping factor from
\citet{Nag11} is in between dashed and solid curves. As an example, we show
their measurements for relaxed clusters in CSF simulations with dotted
curve in Fig. \ref{fig:clfac}.

 The clumping factor, calculated using eq.~\ref{eq:clfac}, directly
  characterizes the increase of the surface brightness in the 0.5-2
  keV band due to overall ellipticity and inhomogeneities in the ICM. Inspection of the mock
  images suggests that very bright and localized regions are
  responsible for high values of $C_X$ when the bulk and high density
  components are considered together (dashed line in
  Fig.~\ref{fig:clfac}). This means that simple cleaning of X-ray
  images from the most obvious bright spots should bring the
  clumpiness factor down to the values characteristic for the bulk
  component (solid line in Fig.~\ref{fig:clfac}).

\begin{table}
 \centering
  \caption{Clumping factor calculated using eq. (\ref{eq:clfac}) for
    total gas from simulations and bulk gas only at $r_{500}$ and
    $1.5~r_{500}$. See also Fig. \ref{fig:clfac}. For comparison, we
    show clumping factors in PKS 0745-191 \citep{Wal12} and Perseus
    \citep{Sim11} clusters
    calculated from the observed overestimation factor of the gas density.}
  \begin{tabular}{@{}ccc@{}}
  \hline
gas component&$C_X$&$C_X$\\
 &  at $r_{500}$ & at $1.5~r_{500}$ \\             
\hline
{\bf CSF REL} & &\\
Total&1.2&1.6\\
Bulk&1.1&1.3\\
\hline
{\bf NR REL} & &\\
Total &1.6&3.1\\
Bulk &1.2&1.4\\
\hline
{\bf CSF UNREL} & &\\
Total &1.6&1.9\\
Bulk &1.4&1.6\\
\hline
{\bf NR UNREL} & &\\
Total &2.2&2.7\\
Bulk &1.7&1.8\\
\hline
PKS 0745-191&$\sim$ 1-3&$\sim$ 2-9\\
\hline
Perseus& $\sim$ 1-3& $\sim$ 9-12\\
\hline
\label{tab:clumping}
\end{tabular}
\end{table} 

For hot
  clusters the temperature dependence of the emissivity is usually
  neglected and the increase of the surface brightness directly
  translates into the overestimation of the density by a factor $\disp
  \sqrt{C_X}$. This quantity is especially important when the
  resulting density profile is used to infer the gas mass or the ratio
  of the gas mass to the total mass, i.e. $f_{gas}$. From this point
  of view it is interesting to compare the results of the simulations
  with the suggested overestimation factor of the gas density from the
  X-ray observations. For the Perseus cluster this factor is $\sim$ 1-1.6
  at $r_{500}$ and increases to $\sim$ 3-3.4 by $1.5~r_{500}$
  \citep{Sim11}. Corresponding value in terms of $C_X$ is the square
  of this factor, i.e. $C_X\approx 1-2$ and $\approx 9-12$ at
  these two radii respectively. For the PKS 0745-191 cluster clumping
  factor $C_X\approx 1-3$ and 2-9 at $r_{500}$ and $1.5~r_{500}$
  respectively \citep{Wal12}. Since the most prominent high density
  peaks were excluded from the X-ray images of both clusters, these
  values of the clumping factor should be close to the values for the
  bulk gas in simulations. Table \ref{tab:clumping} shows values of
  the clumping factor from the simulated clusters in our sample at
  $r_{500}$ and $1.5~r_{500}$. One can notice that there is an
  agreement between simulations and observations at $r_{500}$. At
  $1.5~r_{500}$ simulations are marginally consistent with clumping
  factor in the PKS 0745-191 cluster. However, values for the Perseus
  cluster are larger at $1.5~r_{500}$ than found in simulations
  especially if relaxed sample is considered.

The difference between observations and simulations can be due to
several reasons. The high clumping factor in the Perseus cluster was
inferred from a narrow stripe along the NW arm of the cluster.  Hydrodynamical simulations indicate
that the distribution of gas clumps is highly anisotropic, azimuthal
scatter of the clumping factor is large \citep{Eck12}. It is therefore  possible
that the clumping factor along the given direction is larger
than the azimuthally averaged value. We note that clumping factor for the
PKS 0745-191 is averaged over five directions and is in a better
agreement with simulations.   Another possible
reason of the higher clumping factor in the X-ray data is the contamination by unresolved sources
in low surface brightness regions. The spatial resolution of Suzaku is
limited and it is difficult to properly model the contribution of
background (AGNs) which could lead to contamination and hence the
over-estimate of the gas density in low surface brightness regions in
cluster outskirts. Indeed, it was shown recently that the clumping
factor measured in simulations is sufficient for describing the excess
of the gas density measured with ROSAT observatory in cluster
outskirts \citep{Eck12}. Clearly, more work from both simulations and
observations is needed to resolve the issues discussed above.    

\section{Origin of the density fluctuations in the bulk component}
\label{sec:origin}

Gas density inhomogeneities should cause observable fluctuations of
the surface brightness in the X-ray images of galaxy clusters. For hot
$T>3~{\rm keV}$ gas the X-ray emissivity weakly depends on temperature
or metal abundance. Therefore X-ray surface brightness fluctuations are
a direct proxy for the density inhomogeneities.  Recent analysis of
the surface brightness fluctuations in the Coma cluster shows that the
typical amplitude of density perturbations (RMS) on scales from 30-500
kpc ranges from 5 per cent to 10 per cent \citep{Chu12}. This is in a reasonably
good agreement with the amplitude of density fluctuations we see in
simulations (Fig. \ref{fig:width}) in relaxed clusters. While Coma is
not very relaxed cluster, the central part studied with Chandra and
XMM-Newton is reasonably relaxed. Broad agreement between simulations
and observations is encouraging and suggests that the simulations
might correctly capturing the physics responsible for the density
fluctuations. We now proceed with the discussion of the key properties
of the density inhomogeneities in the simulations.

In the majority of radial shells the density distribution has two distinct
components: (i) log-normal distribution with a substantial width, and
(ii) a high density tail. The latter component is often associated with
the cold and dense gas in subhalos. This component is more
prominent in the CSF simulations. This is not surprising given that the
radiative cooling time of the dense gas can be short.

The width of the log-normal ``bulk'' component is substantial (see
Fig. \ref{fig:sketch} and \ref{fig:width})
-- the mean value of $W_{10}$ at $r_{500}$ varies between 0.15 and 0.5 in most of the
relaxed cluster.  Clearly, the
width is expected to be zero if the potential is spherical and static,
the gas is in equilibrium and radial shells are infinitely narrow. An
interesting question is to understand the properties of the density
fluctuations in the bulk component that cause broadening of the
density distributions.  Below we analyze the properties of these
fluctuations.

\subsection{Finite thickness of radial shells}

We first address the question if the observed spread of densities in a
shell is a spurious effect of the shell finite thickness. Assuming that locally
the number density is a power law function of radius $n_e(r)\propto
r^{-\alpha}$, the upper limit on the total width (from the minimal to
the maximal value) is \be \disp W_{tot}=\disp
log_{10}\left(\frac{n(r)}{n(r+\disp\Delta r)}\right)\approx
\disp\frac{\alpha \frac{\Delta r}{r}}{\ln 10}.  \ee In our
calculations $\disp\frac{\Delta r}{r}\approx 0.01$ and the slope
$\alpha$ of the gas density varies from $\approx 1$ in the centre to
$\approx 3$ at $2~r_{500}$. Therefore, in this case an upper limit on
$W_{10}<W_{tot}$ due to the finite thickness of shells is $\sim 1$ per
cent. We conclude that no significant contribution to the width of
the density distribution is caused by the shell finite thickness.

\subsection{Ellipticity and perturbations of the potential}
\label{sec:pot}

Another plausible reason for the observed variations of the gas
density in a spherical shell is the overall ellipticity/asphericity of
the cluster. One can use the known gravitational potential of the cluster,
created by dark matter and baryons, as a proxy to the underlying
ellipticity. For instance, one can imagine a situation when
essentially hydrostatic gas is sitting in an elliptical potential
well.

For the isothermal gas in hydrostatic equilibrium the number
density $n_e$ is related to the static potential $\phi$ through the Boltzmann
distribution $n_e\propto e^{ -\frac{\phi \mu m_p}{kT}}$.  Let us
calculate a correlation coefficient $C$ between density $\disp\delta \ln
n_e$ 
 and potential $\disp\delta \phi$ variations in a shell. We
define the correlation coefficient $C(x,y)$ between two variables $x$
 and $y$ in a usual way: 
\be 
C(x,y)=C(y,x)=\frac{\langle xy\rangle}{\sigma_x \sigma_y}=\frac{\sum\limits_i
  x_iy_i}{\sqrt{\sum\limits_i x_ix_i}\sqrt{\sum\limits_i y_iy_i}},
\label{eq:corrfunc}
\ee 
where $\sigma_x$ and $\sigma_y$ are the dispersions of $x$ and $y$
respectively; both variables are assumed to have zero mean; $\langle
\rangle$ denotes averaging over all particles in the shell.
\begin{figure}
\begin{minipage}{0.23\textwidth}
\includegraphics[trim=1cm 2cm 1cm 3cm,width=1\textwidth,clip=t,angle=0.]{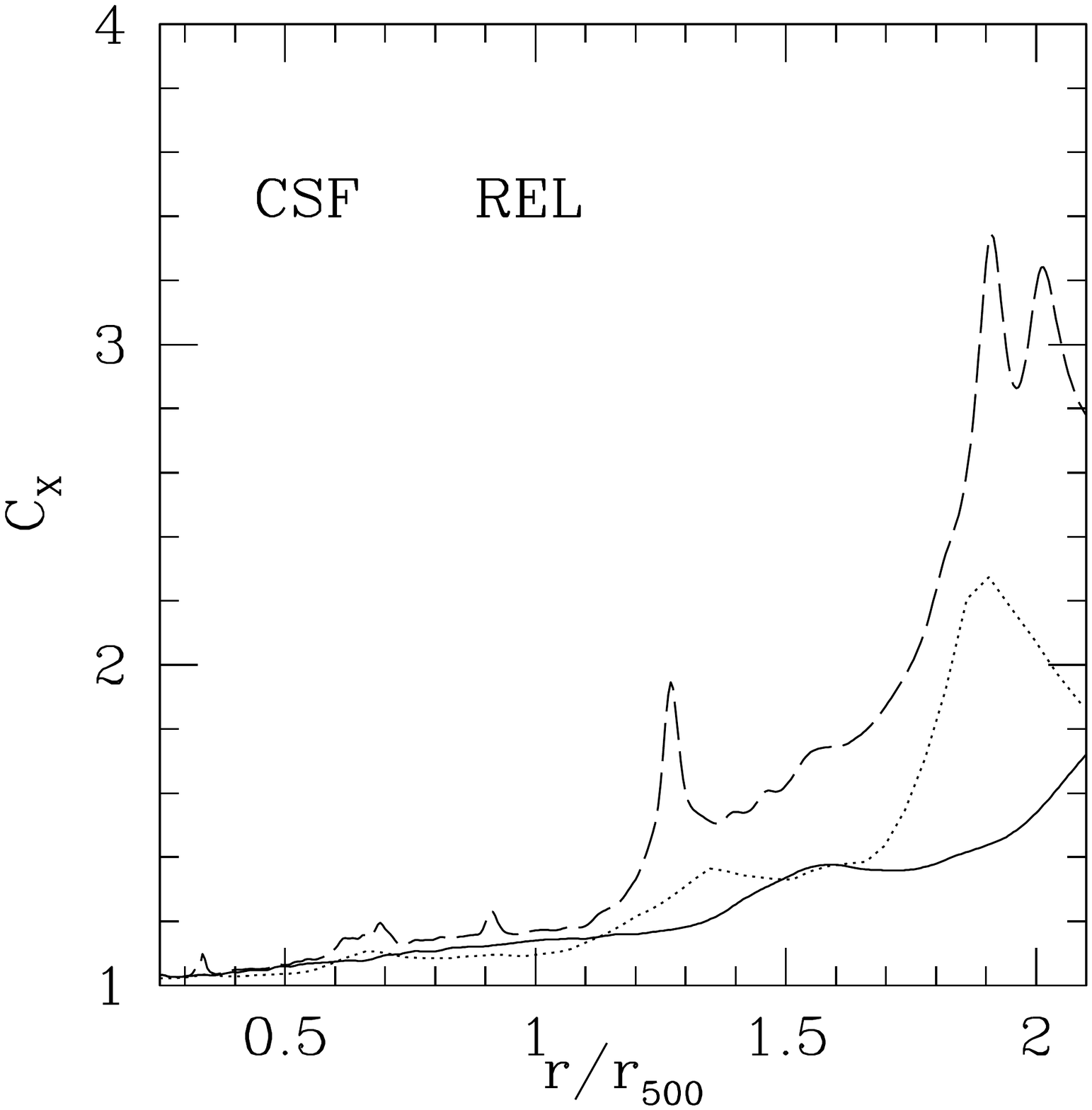}
\end{minipage}
\begin{minipage}{0.23\textwidth}
\includegraphics[trim=1cm 2cm 1cm 3cm,width=1\textwidth,clip=t,angle=0.]{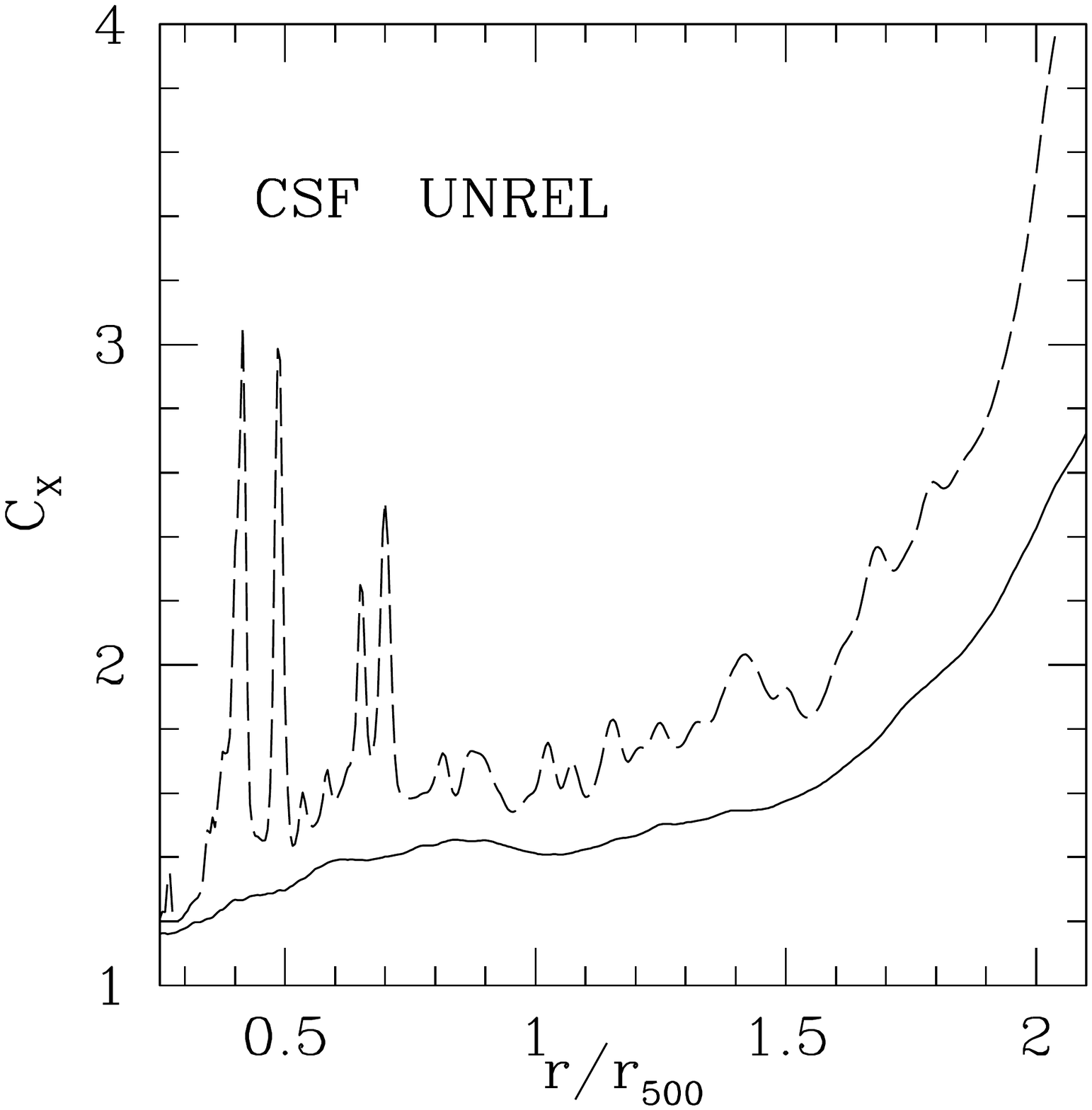}
\end{minipage}\\
\begin{minipage}{0.23\textwidth}
\includegraphics[trim=1cm 4cm 1cm 6cm,width=1\textwidth,clip=t,angle=0.]{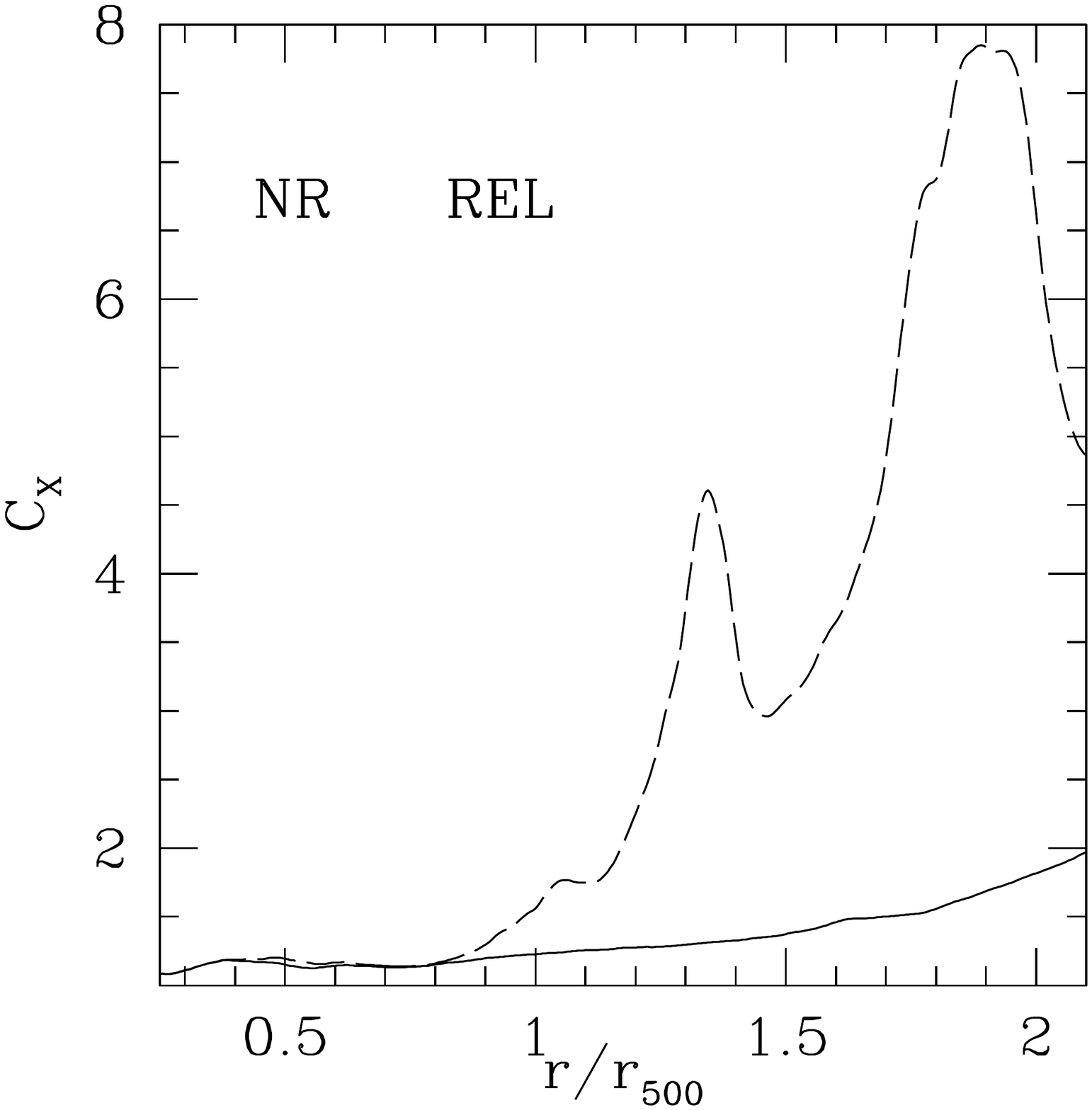}
\end{minipage}
\begin{minipage}{0.23\textwidth}
\includegraphics[trim=1cm 4cm 1cm 6cm,width=1\textwidth,clip=t,angle=0.]{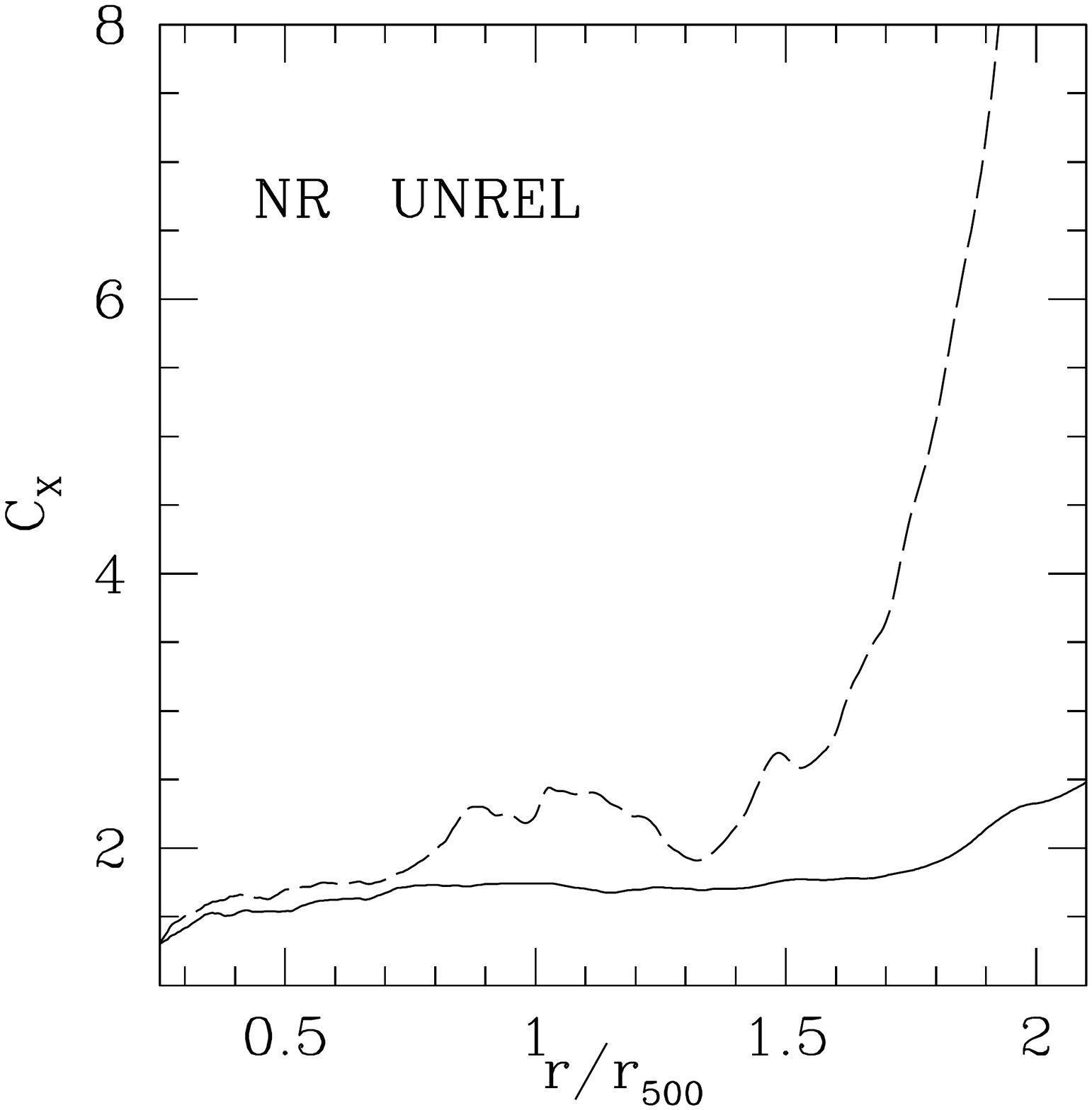}
\end{minipage}
\caption{Clumping factor (eq. \ref{eq:clfac}) as a function of radius,
  averaged over our samples of relaxed/unrelaxed clusters. {\bf Dashed:} full gas
  distribution (bulk plus inhomogeneities), {\bf solid:} bulk
  component. {\bf Dotted:} clumping factor from \citet{Nag11}. See Section
  \ref{sec:cl_factor} for details.
\label{fig:clfac}
}
\end{figure}

\begin{figure}
\begin{minipage}{0.23\textwidth}
\includegraphics[trim=1cm 2cm 1cm 3cm,width=1\textwidth,clip=t,angle=0.]{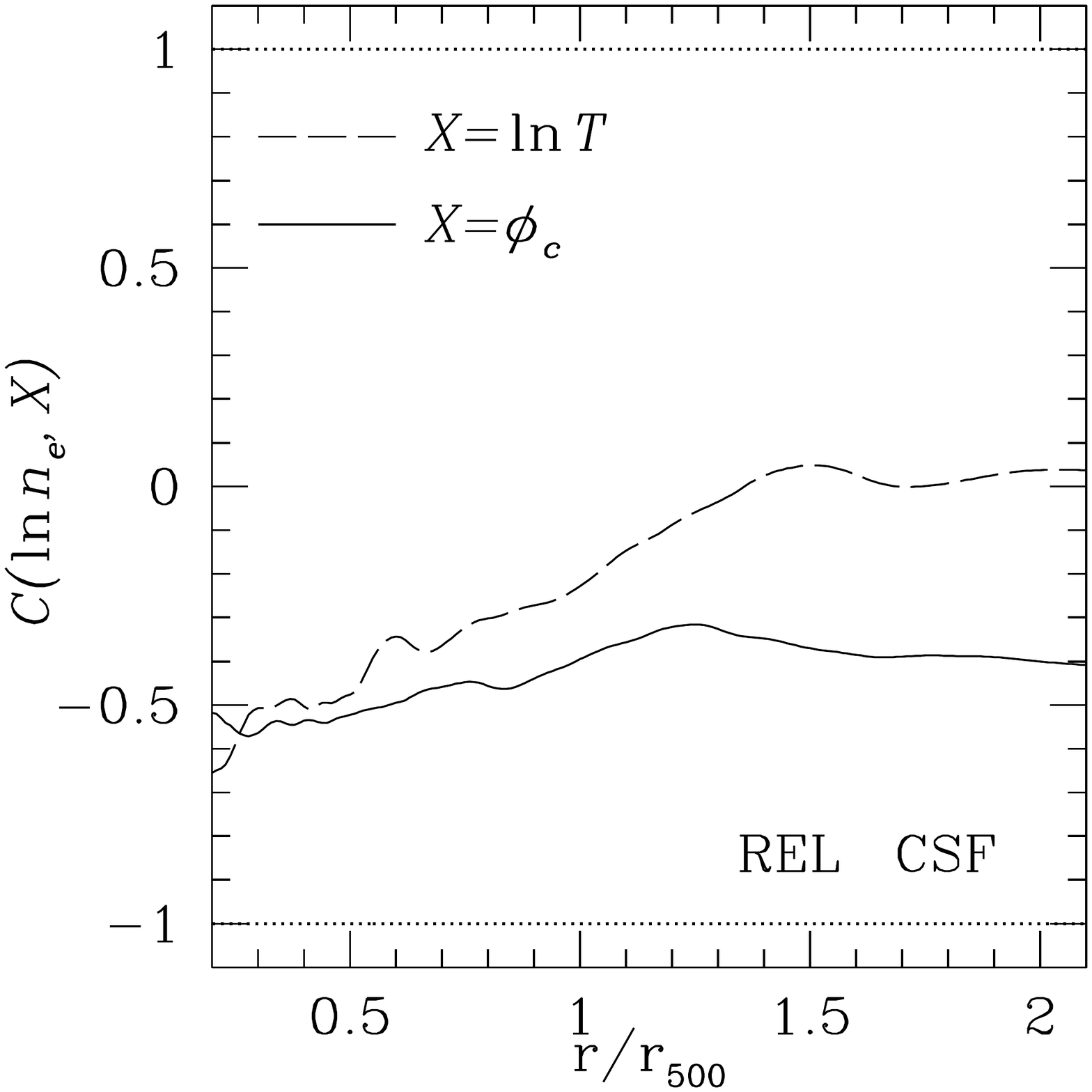}
\end{minipage}
\begin{minipage}{0.23\textwidth}
\includegraphics[trim=1cm 2cm 1cm 3cm,width=1\textwidth,clip=t,angle=0.]{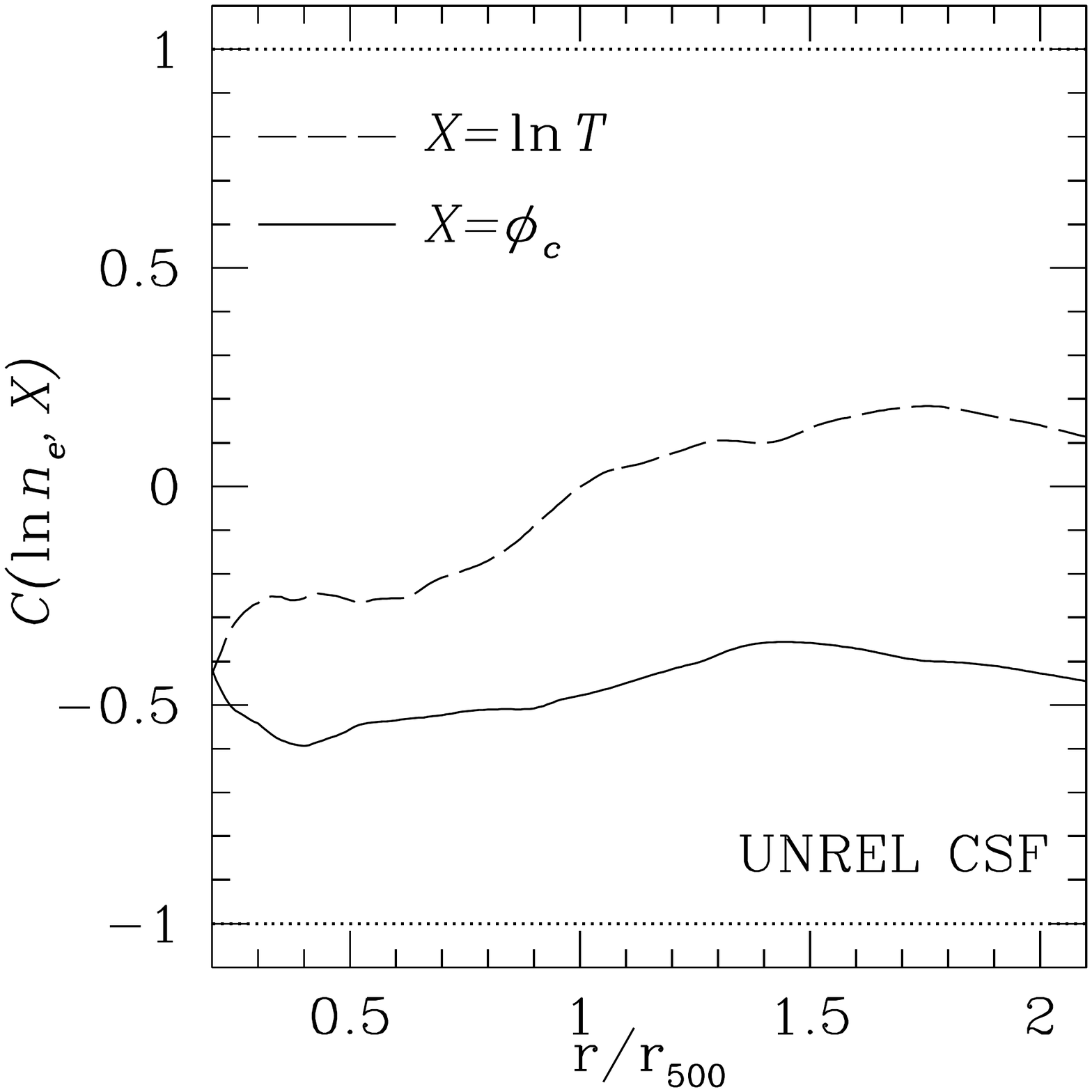}
\end{minipage}\\
\begin{minipage}{0.23\textwidth}
\includegraphics[trim=1cm 4cm 1cm 6cm,width=1\textwidth,clip=t,angle=0.]{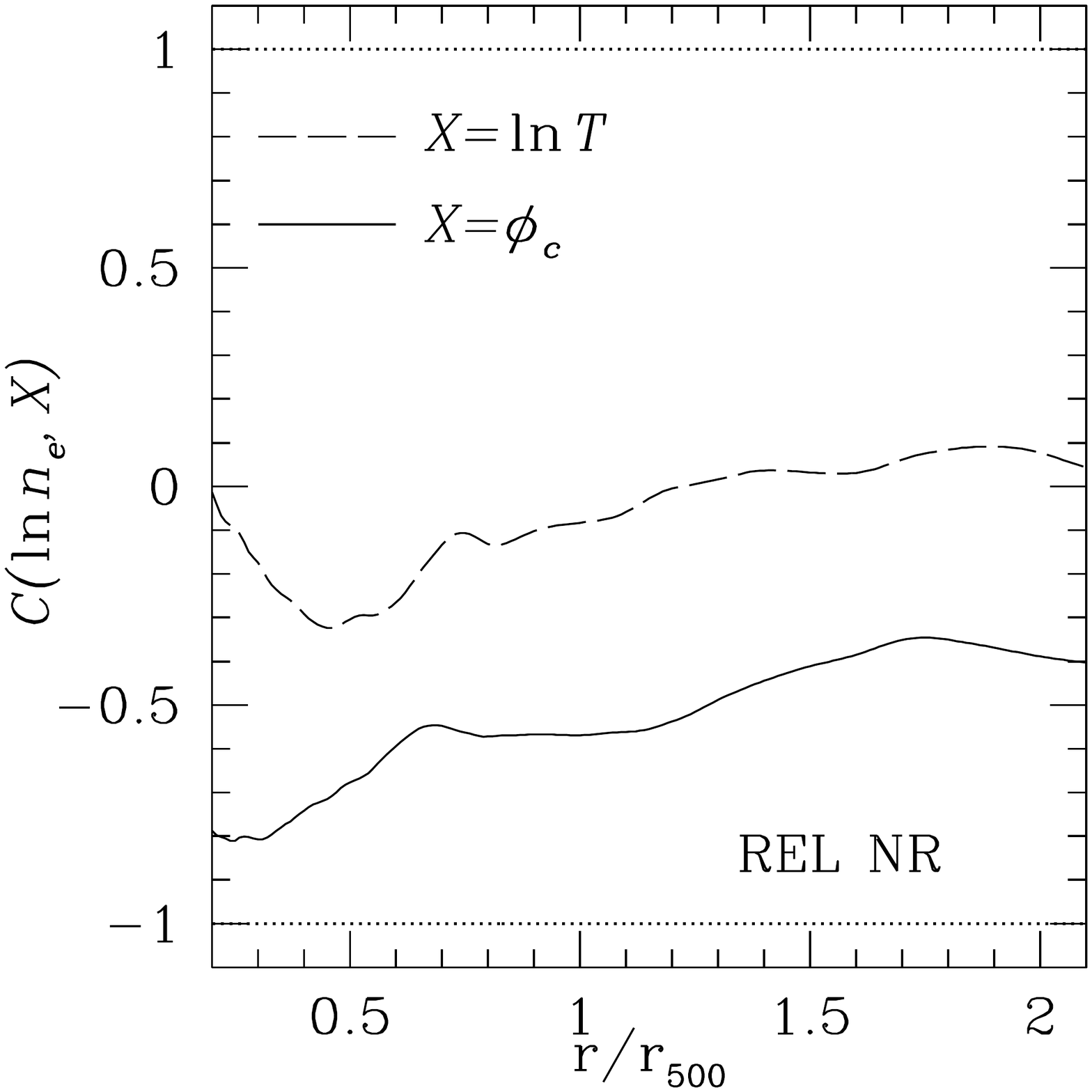}
\end{minipage}
\begin{minipage}{0.23\textwidth}
\includegraphics[trim=1cm 4cm 1cm 6cm,width=1\textwidth,clip=t,angle=0.]{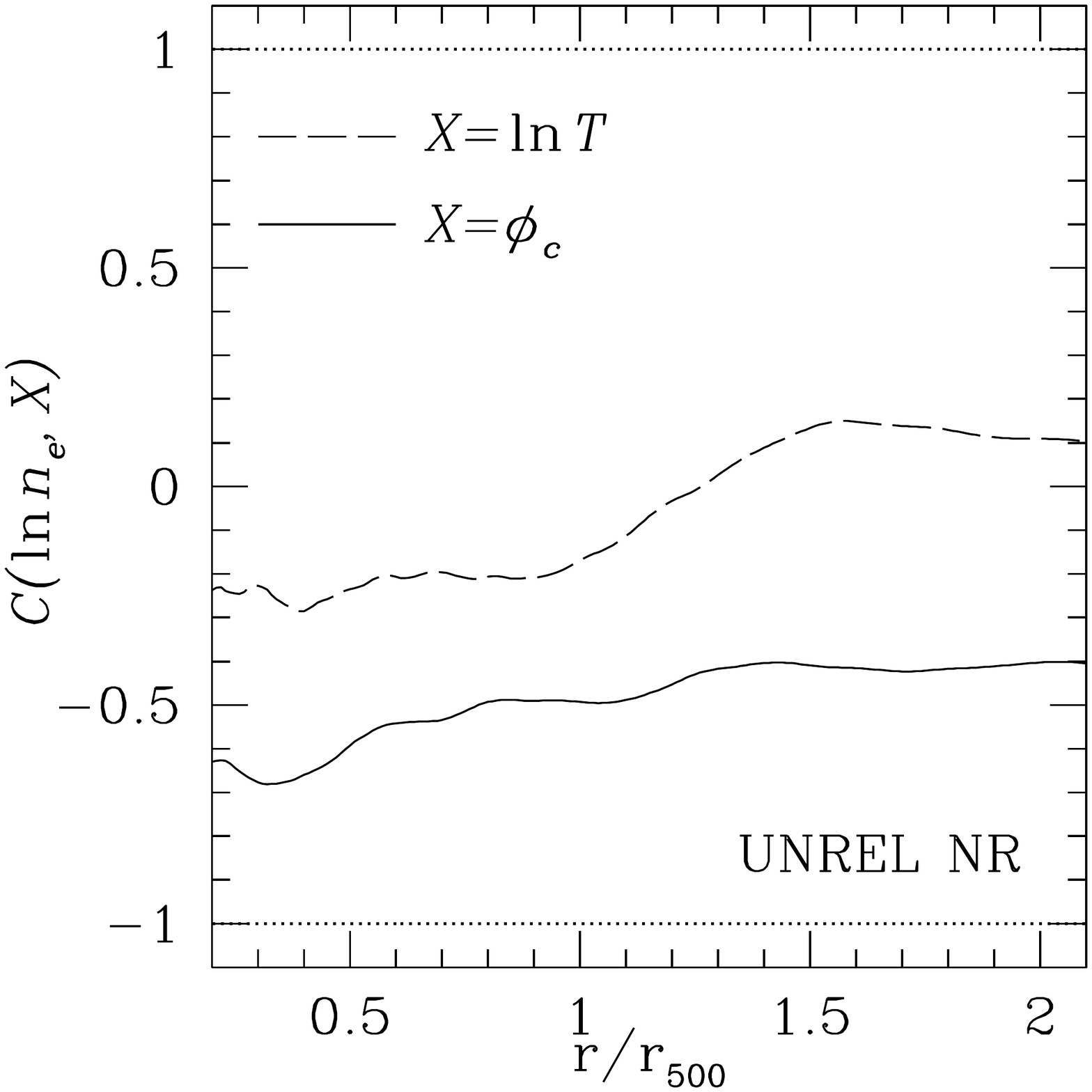}
\end{minipage}
\caption{ Radial profiles of the correlation coefficient
  (eq. \ref{eq:corrfunc}) between the density fluctuations
  $ln(n_e)$ and fluctuations of potential $\phi_c=\phi\disp\frac{\mu m_p}{k
  T_{med}}$ (solid curves) or
  temperature $ln(T)$ (dashed curves). The data are averaged over subsamples of NR/CSF and
  relaxed/unrelaxed clusters. See Sections
  \ref{sec:pot} and \ref{sec:adiiso} for details.
\label{fig:corrcoef}
}
\end{figure}

Fig. \ref{fig:corrcoef} shows the correlation coefficient
$\displaystyle C(\ln n_e,\phi_c)$ (hereafter
$\phi_c=\phi\disp\frac{\mu m_p}{k T_{med}}$) between density and potential
fluctuations for the CF and NR runs averaged over samples of relaxed
and unrelaxed clusters. Clearly, if the isothermal gas is in
equilibrium in a static gravitational potential, then $\disp \delta \ln
n_e=-\phi\frac{\mu m_p}{kT_{med}}$ and the correlation coefficient is
$-1$. In the opposite case $C=1$. However, we see that in simulations the correlation coefficient in
most cases is between $\approx -0.6$ and $\approx -0.4$, except for
relaxed clusters in the NR simulations, where the correlation
coefficient reaches $\approx -0.86$ in the central $r \le
0.5r_{500}$. This means that the asphericity of the potential can not
alone at a
given moment of time explain all observed density
variations. For instance, non-isothermality of the gas in shells could
contribute to the density variations.

Knowing the correlation coefficient between the density and potential one
can calculate the width of the density distribution $W_{10,\phi}$
corrected for the potential variations in the shell. 
Indeed, let us assume that we want to account for correlation
  between variables $x$ and $y$ and construct new variable
  $y'=y-R(x,y)\times x$ with minimal dispersion. We seek the regression coefficient $R(x,y)$ 
  which minimizes $\langle y'^2\rangle$. This regression
    coefficient is (assuming again that both variables have zero mean)
\be
  \displaystyle R(x,y)=\frac{\sum x_iy_i}{\sum x_ix_i}.
\label{eq:regrcoef}
\ee
 The dispersion of $y'$ is then
  $\sigma^2_{y'}=\sigma^2_y-C(x,y)^2\sigma^2_y$. Thus, one can calculate the width of the density
distribution $W_{10,\phi}$ through the
correlation coefficient between the density and potential as
\begin{equation}
W_{10,\phi}=W_{10}\sqrt{1-C(\ln n_e,\phi_c)^2}.  
\end{equation} 
Therefore, from the value of the correlation function $|C(\ln
n_e,\phi_c)|=0.4 - 0.6$ it follows that accounting for the potential
variations in a shell reduces the width of the log-normal density
distribution by a factor $\disp
\frac{W_{10}-W_{10,\phi}}{W_{10}}=1-\sqrt{1-C^2}\approx 8 -  20 \%$.

Overall ellipticity of the mass distribution is not the
only reason for potential variations in a spherical shell. The
variations can also be caused by the presence of subhalos. However the cores of the most prominent and gas rich subhalos have
been excluded as high density inhomogeneities, leaving only outer regions of
the subhalos as a possible contributor to the bulk component density
variations. The above estimate includes both types of variations and
can be used as an upper limit on the variations induced by the
ellipticity.

The bottom line of this exercise is that in the simulations only a small
part of the density variations in spherical shells can be attributed to
the ellipticity/asphericity of the underlying potential, under
the assumption that it is static.

\subsection{Adiabatic and isobaric fluctuations}
\label{sec:adiiso}
We now address the question if the observed density variations in the
bulk gas component are predominantly adiabatic or isobaric. Adiabatic
fluctuations arise from sound waves or weak shocks (and can be associated with the variations of the potential with a shell), while isobaric
fluctuations naturally appear when gases with different entropies are
brought to contact e.g. by ram pressure stripping or turbulent gas
motions.

\begin{figure}
\includegraphics[trim=0.7cm 5.5cm 0.7cm 3.2cm,width=0.5\textwidth,clip=t,angle=0.]{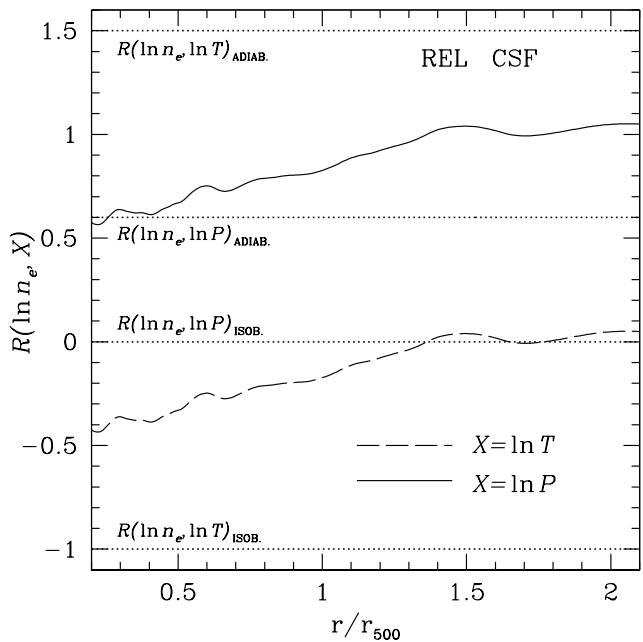}
\caption{Radial profiles of the regression coefficient
  (eq. \ref{eq:regrcoef}) between the density fluctuations and
  temperature or pressure fluctuations. The profiles are averaged over a
  sample of relaxed CSF clusters. Dotted lines show the regression
  coefficient in cases of pure adiabatic and pure isobaric
  fluctuation. The sample-averaged regression coefficients
  do not show the values characteristic for pure adiabatic or pure
  isobaric fluctuations of density. See Section \ref{sec:adiiso} for details.
\label{fig:regrcoef}
}
\end{figure}

 To answer this question we calculate the regression coefficient $R$
defined by eq. \ref{eq:regrcoef} between density variations and
temperature or pressure variations (Fig. \ref{fig:regrcoef}). If density fluctuations are purely adiabatic,
then the regression coefficient $R(\ln n_e,\ln T)=1/(\gamma -1)=1.5$
and $R(\ln n_e,\ln P)=1/\gamma=0.6$. In the case of pure isobaric
density fluctuations, $R(\ln n_e,\ln T)=-1$ and, obviously,
$R(\ln n_e,\ln P)=0$. One can see in Fig. \ref{fig:regrcoef} that the sample-averaged (over sample of relaxed CSF clusters)
regression
coefficient does not correspond to the value characteristic for pure adiabatic
or isobaric density fluctuations. This conclusion remains valid for unrelaxed
clusters and for NR runs as well. In individual clusters the scatter in the
regression (and correlation) coefficients is large. The value of $R(\ln n_e,\ln T)$  varies
from -0.8 up to 0.5, i.e. in some cases $R$ approaches values
characteristic for pure isobaric fluctuations. Inspection of individual clusters
shows that these low/high values of $\disp C(\ln n_e, \ln T)$ at some
radii are often driven by some distinct feature, like an outskirt of a
subhalo or a moderately strong shock.

\subsection{General comments on the density variations}
As we discussed above the asphericity of the gravitational potential
at a given moment can account up to 20 per cent of the bulk
component density variations in narrow radial shells. 

Another likely reason for the density variations is directly related
to gas motions. From Fig.~\ref{fig:vmean} it is clear that the gas
motions in the bulk component are predominantly subsonic. We are
therefore dealing with a weakly compressible case. Any given velocity
field can be decomposed into solenoidal and compressible parts, both
of which can contribute to the observed variations of density and
pressure. A crude estimate of the density variations caused by
variations of the square of the gas velocity is possible
using Bernoulli's equation: $\disp \frac{\delta P}{P}\propto M^2$,
where $M$ is a characteristic Mach number of isotropic turbulent gas
motions $\disp M^2=\frac{\langle v_x^2+v_y^2+v_z^2 \rangle}{c_s^2}$
\citep{Lan59}. Assuming that the density fluctuations are adiabatic,
i.e. $\disp \frac{\delta n}{n} \propto \frac{1}{\gamma}\frac{\delta
  P}{P}$ one can expect $\disp \frac{\delta n}{n}\approx
\frac{\alpha}{3}M^2$, where $\alpha$ is of order unity. For an order
of magnitude estimate one can adopt the value of $\alpha\approx 0.7$,
calculated for incompressible gas \citep[e.g.][]{Hin75}. The Mach
number evaluated for the bulk flow is $\sim 0.55$ at $r_{500}$ for
relaxed clusters in both CSF and NR simulations (see
Fig. \ref{fig:vmean}). Therefore\footnote{Recall that numerically
  $W_{10}$ is close to the standard deviation of the log-normal
  distribution.}, $\disp W_{10}\approx\frac{\delta n}{n}\approx 0.07$
at $r_{500}$.  This means that variations of the square of the gas
velocity can explain $\sim 30$ per cent of the width of density
distribution found in simulations. 

The contribution of pure compressible motions to the density
variations scales linearly with the Mach number $M$. To evaluate it
properly one has to make Helmholtz decomposition of the velocity
field. This is beyond the scope of this paper. We instead note that
both solenoidal and compressible modes in the subsonic case should
lead to adiabatic relation between density and pressure (or
temperature) fluctuations. As we saw above (\S\ref{sec:adiiso}) the
mean density/temperature regression coefficient (see
Fig.\ref{fig:corrcoef}) is not close to 1.5. We further estimated the
mean correlation coefficient $\disp C(\ln n_e, \ln P)\sim 0.7$ between
the density and pressure fluctuations at $r_{500}$ for a sample of
relaxed clusters. This corresponds to $\sim$ 30 per cent of the
observed density
variations, placing constraints on both solenoidal and compressible
modes together. Therefore pure adiabatic fluctuations alone are not able to
explain the density variations found in simulations and substantial
contribution should be associated with the entropy variations.

Indeed, time variations of the potential and gas motions can also be
responsible for inhomogeneity of the gas in radial shells, when the
gas  with an entropy different from the
mean/median gas entropy at a given radius is advected to this
radius. One can identify two flavors of this process. First, the motion
of a subhalo can be responsible for the transportation of the gas. This
process is also accompanied by a ram pressure stripping and
partial mixing of the gas with the ICM. Second, the gas motions
themselves can displace
lumps of gas with different entropies from their ``equilibrium''
radius. The morphology of a moderately overdense gas component,
corresponding to 2-3 $\sigma$ deviations from the median value (see
\S\ref{sec:exclusion}) traces the distribution of subhalos,
suggesting the first mechanism as a primary contributor. At smaller
density contrasts $\le 1~\sigma$ it is difficult to draw firm
conclusion. Most likely both mechanisms play a role in producing
nonuniform density in radial shells. Inspection of
Fig. \ref{fig:corrcoef} suggests that at least in the central region
$r<r_{500}$ apparent anti-correlation of the density and temperature
variations is driven by these mechanisms.

\section{Sensitivity of results to physics included in simulations}
\label{sec:sens}
Our analysis was applied to simulated galaxy clusters with different
physics included. We expect the real characteristics of the ICM
gas to be somewhere in between CSF and NR cases. It is therefore
interesting to examine which quantities, studied here, changes
strongly between CSF and NR runs.

In general, gas cooling in CFS runs should lead to a more dense and
prominent core at the cluster centre. The off-centre clumps are also
expected to be denser and more compact. These effects reveal
themselves as a ``forest'' of high density peaks in the density
distribution in the CFS runs (see Fig.~\ref{fig:cl7_distr}). The
impact of the cooling on the bulk component is much more subtle, as
summarized below for a subsample of relaxed clusters:
\begin{itemize}
\item The difference between the mean width of density and pressure
  distributions in the CSF and NR runs is minor. At $r_{500}$ the
  mean width of density distributions in relaxed clusters is $\sim$ 16 per cent larger in NR runs than in the CSF
  ones. The difference in the width could be due to the difference in
  ellipticity of the gas distribution
  \citep[see][]{Lau11,Lau12}. Indeed, after correction for the
  contribution of the ellipticity, using the approach, outlined in
  \S\ref{sec:pot}, the difference in the width of the distributions in
  the NR and CSF runs near $r_{500}$ reduces  down to $\sim$4 per cent. 
\item The RMS velocities in the bulk component or the ratio $\disp
  \frac{P_{motions}}{P_{thermal}}$ are very similar in both runs.
\item The clumping factor (see eq. \ref{eq:clfac}), calculated for the
  bulk component at $r_{500}$ is $\sim 23$ per cent and $\sim 8$ per
  cent higher in the NR
  runs for unrelaxed and relaxed clusters respectively. The clumping factor, calculated for total density distribution
  including the high density tail is also larger for NR run. In this
  case the difference can be up to 40 per cent. The clumping factor is
  sensitive to cluster classification on relaxed and unrelaxed systems
  and has a very large scatter from cluster to cluster. Even after
  averaging over the sample, one object can dominate the mean.
\end{itemize}
This difference between CSF and NR can be understood in terms of a
simple notion: in CSF runs the gas with high or intermediate densities
has a short cooling time. This gas cools down below X-ray temperatures
resulting in a stronger separation of hot and low density gas and much
colder clumps. While in the NR runs the gas at intermediate
densities/temperatures has much longer life time.

\section{Summary}
\label{sec:sum}

In this study we propose a novel description of the ICM in simulated
galaxy clusters that allows us to better understand the properties of the
bulk of the hot gas in clusters and various inhomogeneities. Our
analysis is applied to 16 simulated galaxy clusters with different
baryonic physics. The main results and conclusions can be summarized
as follows.

{\bf 1.} We suggest a simple, quick and robust method to divide the
ICM in simulations into a nearly hydrostatic bulk component and
non-hydrostatic high density inhomogeneities. This allows us to study
separately the properties of both components. In X-ray observations
similar division between these two components is usually based on the
analysis of X-ray images from which bright localized spots are
excluded. The selection of the bulk component implemented in the
present study corresponds to the idealized case, when statistical
quality of the data allows one to make careful cleaning of the image
from all distinct features. The analysis of two components together
corresponds to another limit when no bright spots are excluded from
observed images.

{\bf 2.} The characteristic amplitude of stochastic gas
velocities in the bulk component is increasing with radius and has a
very regular behavior. RMS velocity averaged over a sample of relaxed
(unrelaxed) clusters varies from $\sim 0.4~c_{s,500} (\sim 0.7~c_{s,500})$ at
$0.3~r_{500}$ to $\sim 0.6~c_{s,500} (\sim 0.8~c_{s,500})$ at
$2~r_{500}$. This is in a broad agreement with previous measurements
\citep[e.g.][]{Lau09}. Velocities of motions in the high density component, in
contrast, are higher, change irregularly with radius and exhibit
large scatter from one cluster to another. Note that the values of
velocities in the bulk gas component are consistent with current
velocity estimates from X-ray observations
\citep[e.g.][]{Wer09,San11,Pla12}. The forthcoming {\it
  Astro-H}\footnote{http://astro-h.isas.jaxa.jp/}  mission (launch
date 2014) will provide more robust constraints on the gas
velocities since it has high energy resolution $\sim 5$
  eV. E.g. in \citet{Zhu12} we show that in the Perseus-like clusters
  gas motions with  $V_{gas}\gtrsim 0.3 c_s$ cause line broadening of
  $\sim$ 7 eV, while the thermal broadening does not exceed 3
  eV. Therefore, gas motions should be readily detectable.

A closely related quantity is
the ratio of pressure due to gas motions and thermal pressure, which is
$\sim 0.2$ at $r_{500}$ for relaxed clusters. This result holds for
bulk component alone and for the bulk plus high density components
together. At larger radii the ratio $\disp
\frac{P_{motions}}{P_{thermal}}$ in the bulk component is increasing
gradually to 50-60 per cent at $2~r_{500}$ for relaxed clusters (and even more for the bulk
plus high density components together).

{\bf 3.} The clumping factor boosts the X-ray emissivity for the bulk
component in relaxed clusters by less than $\sim$15-25 per cent within
$r_{500}$. This factor increases to 30-40 per cent at $2~r_{500}$. For the bulk
plus high density components together the clumping factor is much
larger and is very irregular.

{\bf 4.} We introduce two characteristics of the bulk component:
median radial profiles of density/temperature/pressure -
characteristic of the overall radial properties and width $W_{10}$ of
the density and pressure distributions - characteristic of gas
fluctuations around the median value.

In contrast to the mean, mode or range profiles, the median profiles
are very robust even if the ICM is strongly contaminated by high
density inhomogeneities. Therefore, in order to calculate radial
profiles of gas characteristics, we do not need to exclude clumps from
the ICM first. This would significantly simplify analysis of big
samples of simulated clusters.

The density distribution of the bulk gas in each radial
shell can be well described by log-normal distribution. We
propose to use the width of density distributions as another robust
characteristic of the bulk gas. The width is an
increasing function of distance from the cluster centre with a
relatively small scatter from cluster to cluster, especially for
relaxed clusters. The typical width of the density distribution at
$r_{500}$ in our CSF sample is $W_{10}\sim 0.25$ dex with the scatter $\pm 0.037$
for relaxed clusters, while for unrelaxed clusters the typical width is
$\sim 0.43$ dex with slightly larger scatter $\pm 0.14$.  This
  suggests that the width can be used as an additional criterion to
classify clusters in large simulations into relaxed and unrelaxed
clusters.

{\bf 5.} We investigated the properties of the density
  inhomogeneities in the simulated sample. The ellipticity of the
  underlying mass distribution can explain $8-20$ per cent of the
  observed density variations of the bulk component in individual
  clusters at $r_{500}$. Another $\sim 30$ per cent of the density
  distribution width $W_{10}$ at $r_{500}$ can be attributed to the
  adiabatic pressure/density variations in the turbulent ICM. The
  remaining part of the observed density variations in the bulk
  component is associated with the variations of gas entropy at a
  given distance from the cluster centre. These entropy variations are
  likely caused by advection of the gas by moving subhalos, including
  the ram-pressure stripped gas,
  and by gas advection by stochastic gas motions.

{\bf 6.} Compared to observations, the width of the gas density
  distribution in the inner parts of relaxed clusters $W_{10}\sim
  0.1-0.2$ dex (FWHM), broadly agrees with the typical amplitude of density
  perturbations of 5 per cent to 10 per cent (RMS) in the Coma cluster core
  \citep{Chu12}. In the cool-core AWM4 cluster, which is probably more
  relaxed than Coma, \citet{2012MNRAS.421..726S} found 4 per cent
  density variations. Further analysis of a sample of cluster is
  needed to conclude if additional processes like thermal conduction
  or mixing are required to reduce the ICM inhomogeneity in 
  simulations. The clumping factor $C_X \sim 1.1-2.2$ at $r_{500}$
    found in the simulations is in agreement with the value 1-3
    suggested by the observations. However, at cluster outskirts ($1.5
    ~r_{500}$) clumping from the simulations broadly agrees with
    observations of PKS0745-191 cluster \citep{Wal12} but is $\sim 3$
    times smaller than the clumping factor in the Perseus cluster\citep{Sim11}.

While this paper was in review, we noticed another paper by
\citet{Bat12}, where the measurement biases of $f_{gas}$ were analyzed using another sample of simulated clusters. In
their paper \citet{Bat12} used $T>10^6$K cut in the calculation of the
clumping factor to mimic the X-ray observations
(cf. eq. \ref{eq:clfac}) and provide the value of $f_{gas}$ within
given radius. Crude estimates show that the behavior of the clumping factor is qualitatively similar to our results for the case when high density inhomogeneities are not excluded (dashed line in Fig.\ref{fig:clfac}).

{\bf 7.} The analysis described in the paper has important
implications for the measurements of the total mass of
clusters. Separation of the high density tail from the bulk component
of the ICM, allows one to plug into HSE equation quantities that are
not contaminated by various gas inhomogeneities. This provides a lower
limit on the mass bias one can obtain from the standard analysis of
X-ray observations, once all substructures are carefully removed from
the data. This issue will be addressed in our future work.   Also, a combination of the median density and temperature
radial profiles along with the width of their distributions provides a
convenient way to calculate other possible biases in the observables,
such as the bias in the X-ray emissivity or X-ray temperature and
pressure measured through the SZ effect \citep[see][]{Khe12}.

\section{Acknowledgements} 
 IZ, EC, AK, DN and EL thank KITP for
hospitality during the workshop ``Galaxy Clusters: the Crossroads of
Astrophysics and Cosmology''(2011). The work was supported in part by the Division of
Physical Sciences of the RAS (the program ``Active processes in
galactic and extragalactic objects'', OFN-17). AK was supported in part by NSF grants AST-0807444, NASA grant
NAG5-13274, and by the Kavli Institute for
Cosmological Physics at the University of Chicago through
the NSF grant PHY-0551142 and PHY-1125897 and an endowment from the
Kavli Foundation. EL was supported in part by NASA Chandra Theory grant GO213004B. DN 
acknowledges support from NSF grant AST-1009811, NASA ATP 
grant NNX11AE07G, NASA Chandra Theory grant GO213004B, Research 
Corporation, and Yale University. This work was supported in part by the facilities and staff of
the Yale University Faculty of Arts and Sciences High Performance
Computing Center. The cosmological simulations used in this study were performed on the IBM RS/6000 SP4 system (copper) at the National Center for Supercomputing Applications (NCSA).

\label{lastpage}
\end{document}